\documentclass{article}
\usepackage{spconf,amsmath,graphicx}

\usepackage{enumitem}
\setlist{nosep, leftmargin=14pt}

\usepackage{mwe} % to get dummy images

\usepackage{amsmath,amssymb,amsthm}
\usepackage{graphicx}
\graphicspath{figure/}
\usepackage{subfigure}
\usepackage{caption}
\usepackage{float}
\usepackage{bm}
\usepackage{diagbox}

\usepackage{algorithm}
\usepackage{algorithmicx}
\usepackage{algpseudocode}
\usepackage{url}
\usepackage{indentfirst}
\usepackage{verbatim}
\usepackage{booktabs,multirow}
\usepackage{tikz}
\usepackage{cite}
\usetikzlibrary{spy}
\usepackage{booktabs}
\usepackage{multirow}
\usepackage{stfloats}

\newcommand{\x}		{\mathbf{x}}	
\newcommand{\y}		{\mathbf{y}}
\newcommand{\A}		{\mathbf{A}}

\newcommand{\W}	 {\mathbf{W}}
\renewcommand{\P}	 {\mathbf{P}}
\newcommand{\omg}	 {\mathbf{\Omega}}

\newcommand{\Z}		{\mathbf{Z}}
\newcommand{\0}	     {\mathbf{0}}

\newcommand{\I}	     {\mathbf{I}}

\newcommand{\R}	     {\mathbf{R}}

\newcommand{\U}		{\mathbf{U}}	
\newcommand{\V}		{\mathbf{V}}
\newcommand{\Sig}		{\mathbf{\Sigma}}

\title{TWO-LAYER CLUSTERING-BASED SPARSIFYING TRANSFORM LEARNING FOR LOW-DOSE CT RECONSTRUCTION}
\name{Xikai Yang$^1$, Yong Long$^1$, Saiprasad Ravishankar$^2$ 
}
\address{$^1$University of Michigan - Shanghai Jiao Tong University Joint Institute, \\
	Shanghai Jiao Tong University, Shanghai 200240, China\\
	$^2$Department of Computational Mathematics, Science and Engineering \\ and Department of Biomedical Engineering,
	Michigan State University, East Lansing, MI 48824, USA
	}
\begin{document}
%\ninept
%
\maketitle
\begin{abstract}
Achieving high-quality reconstructions from low-dose computed tomography (LDCT) measurements is of much importance in clinical settings. Model-based image reconstruction methods have been proven to be effective in removing artifacts in LDCT. In this work, we propose an approach to learn a rich two-layer clustering-based sparsifying transform model (MCST2), where image patches and their subsequent feature maps (filter residuals) are clustered into groups with different learned sparsifying filters per group. We investigate a penalized weighted least squares (PWLS) approach for LDCT reconstruction incorporating learned MCST2 priors. Experimental results show the superior performance of the proposed PWLS-MCST2 approach compared to other related recent schemes.
\end{abstract}
\vspace{-0.05in}
\vspace{-0.15in}
\section{Introduction}
\vspace{-0.1in}
Low-dose computed tomography (LDCT) has received much interest in clinical and other settings. A predominant challenge in LDCT is to obtain high-quality reconstructions despite the reduced intensity of radiation. Traditional methods such as analytical filtered back-projection (FBP)~\cite{feldkamp:84:pcb} perform poorly for LDCT reconstruction, and produce substantial streak artifacts. Model-based image reconstruction~\cite{fessler:00:sir} methods have been especially popular for LDCT. In particular, penalized weighted least squares (PWLS) approaches incorporating the edge-perserving (EP) regularizer~\cite{cho:15:rdf} significantly reduce the artifacts present in FBP images.

Other works have proposed an adaptive regularization term for statistical iterative reconstruction. In particular, there has been growing interest in designing data-driven regularizers that capture complex sparse representations of signals from training datasets~\cite{aharon:06:ksa,rubinstein:13:aka}. Sparsifying transform learning~\cite{ravishankar:13:lst} is a generalization of analysis dictionary learning and is an approach for learning models that when applied to images approximately sparsify them. Compared to conventional synthesis dictionary learning methods~\cite{aharon:06:ksa} that are often NP-Hard and involve expensive algorithms, sparsifying transform learning methods are computationally very efficient due to closed-form sparse code and transform updates. In particular, the optimal sparse coefficients in the transform model are typically found by thresholding operations.

Due to their low computational cost, several transform learning-based methods have been studied for image reconstruction in recent years
%~\cite{zheng:18:pua, ravishankar:18:lml}, 
including the union of transforms approach based on data clustering (ULTRA)~\cite{zheng:18:pua} and multi-layer sparsifying transform (MRST) models~\cite{ravishankar:18:lml,yang:20:lmr}, where the transform domain feature maps (filter sparsification residuals) are sequentially sparsified over layers. Although, both ULTRA and learned MRST models offer benefits for LDCT reconstruction, the MRST model tends to oversmooth image details~\cite{yang:20:lmr}. On the other hand, the union of transforms (ULTRA) model can flexibly capture a diversity of image edges and subtle details and contrast by learning a transform for each class of features, which motivates combining its benefits with the richness of deep transform models.

In this paper, we propose \emph{unsupervised} learning of a two-layer clustering-based sparsifying transform model (referred to as MCST2) for images, where image patches and their feature maps (filtering residuals) in the transform domain are clustered into different groups, with a different learned transform per group. The image patches or features in each group are assumed sparse under a common transform. We derive an exact block coordinate descent algorithm for both transform learning and for image reconstruction with a learned MCST2 regularizer. We investigate the performance of PWLS with MCST2 regularization for LDCT reconstruction. Our experimental results show that MCST2 achieves improved image reconstruction quality compared to several recent learned sparsity-based approaches. PWLS-MCST2 also significantly outperforms conventional methods such as FBP and PWLS-EP.
\vspace{-0.2in}

\section{Algorithm for Model Training}
\vspace{-0.15in}
\subsection{Problem formulation}
\vspace{-0.10in}
Our proposed method is patch-based. The underlying cluster optimization over two layers groups the training patches and their corresponding filter residuals (feature maps) into different classes. The transform domain residuals in the second layer contain finer structures that are sparsified with a union (collection) of transforms.
Our formulation for training the MCST2 model is as follows, with $\mathcal{H}_0 = \{\omg_{1,k},\omg_{2,l},\Z_{1,i},\Z_{2,j},C_{1,k},C_{2,l}\}$ denoting the set of \emph{all} optimized variables:
\vspace{-0.15in}
\begin{equation}\label{eq:P0}
\vspace{-0.05in}
\begin{aligned}
&\min_{\mathcal{H}_0} \sum_{k=1}^{K} \sum_{i\in C_{1,k}} \hspace{-0.05in} \|\omg_{1,k}\R_{1,i}\hspace{-0.02in}-\hspace{-0.02in}\Z_{1,i}\|_2^2 \hspace{-0.02in}+\hspace{-0.02in} \eta_1^2\|\Z_{1,i}\|_0 
\\ 
&\quad +  \sum_{l=1}^{L} \sum_{j\in C_{2,l}}  \|\omg_{2,l}\R_{2,j}-\Z_{2,j}\|_2^2 + \eta_2^2\|\Z_{2,j}\|_0 , 
\\
&\mathrm{s.t.} \;\R_{2,i} \hspace{-0.05in}= \omg_{1,k}\R_{1,i}\hspace{-0.05in}-\Z_{1,i}, \forall i \in C_{1,k}, \forall k, \;\;\; \{C_{1,k}\} \hspace{-0.03in}\in \mathcal{G}_1,  
\\
&\quad \{C_{2,l}\}\hspace{-0.03in} \in \mathcal{G}_2, \;\;\;\omg_{1,k}\omg_{1,k}^T  = \omg_{2,l} \omg_{2,l}^T = \I, \;\forall k,l, 
\end{aligned}
\tag{P0}
%\vspace{-0.05in}
\end{equation}
where $\R_1\in \mathbb{R}^{p\times N'}$ denotes the training matrix, whose columns $\R_{1,i}$ represent vectorized patches extracted from images, and $\R_2$ denotes the residual map obtained by subtracting the transformed patches and their sparse approximations. In particular, $\Z_1$ and $\Z_2$ denote the sparse coefficient maps for the two layers, with $\eta_1$ and $\eta_2$ denoting sparsity controlling parameters. We assume that the image patches and residual matrix columns can be grouped into $K$ and $L$ disjoint classes, respectively, with $\{\omg_{1,k}\}$ and $\{\omg_{2,l}\}$ denoting the learned sparsifying transforms in the first and the second layer, respectively. We let $C_{1,k}$ and $C_{2,l}$ denote the sets containing the indices of the columns of $\R_1$ and $\R_2$ that belong to the $k$th class in the first layer and the $l$th class in the second layer, respectively. The sets $\mathcal{G}_1$ and $\mathcal{G}_2$ include all the possible disjoint partitions of $[1:N']$ into $K$ and $L$ sets, respectively.
\vspace{-0.10in}
\subsection{Transform Learning Algorithm}
\vspace{-0.05in}
We propose an exact block coordinate descent (BCD) algorithm to optimize \eqref{eq:P0} by alternatively updating $\{\Z_{1,i}, C_{1,k}\}$, $\{\Z_{2,j}, C_{2,l}\}$, and the transforms $\omg_{1,k}$ and $\omg_{2,l}$. When updating each set of variables, the other variables are kept fixed. Since the solutions to the subproblems are computed exactly, the objective function in \eqref{eq:P0} converges over the BCD iterations. 
\vspace{-0.2in}
\subsubsection{Update Coefficients and Clusters in the First Layer}
\vspace{-0.10in}
Here, we solve \eqref{eq:P0} with respect to the coefficients and cluster memberships in the first layer ($\{\Z_{1,i},C_{1,k}\}$), with the other variables fixed. This leads to the following subproblem \eqref{eq:sub_P0_Z1}, whose exact solution is shown in
\eqref{eq:update_C1} and \eqref{eq:update_Z1} (can be derived similar to~\cite{zheng:18:pua} and \cite{yang:20:mars}), with $H_\eta(\cdot)$ denoting the hard-thresholding operator that sets vector elements with magnitude less than $\eta$ to zero.
\vspace{-0.10in}
%\begin{small}
\begin{equation}\label{eq:sub_P0_Z1}
\begin{aligned}
\min_{\hspace{-0.10in}\big\{\substack{\Z_{1,i},\\C_{1,k}}\big\}} &\sum_{k=1}^{K} \sum_{i \in C_{1,k}} \hspace{-0.10in} \|\omg_{1,k}\R_{1,i}\hspace{-0.05in}-\Z_{1,i}\|_2^2 + \eta_1^2\|\Z_{1,i}\|_0  
\\
& + \sum_{l=1}^{L} \sum_{j\in C_{2,l}}  \|\omg_{2,l}\R_{2,j}-\Z_{2,j}\|_2^2.
\end{aligned}
\vspace{-0.15in}
\end{equation}
%\vspace{-0.10in}
%\begin{equation}\label{eq:update_Z1}
%\hat{\Z}_{1,i}\hspace{-0.05in}=H_{\eta_1/\sqrt{2}}(\omg_{1,k}\R_{1,i}-0.5\omg_{2,l}^T\Z_{2,j}) \quad \forall i \in C_{1,k}.
%\vspace{-0.05in}
%\end{equation}
\vspace{-0.20in}
\begin{equation}\label{eq:update_C1}
\vspace{-0.05in}
\begin{aligned}
\hat{k}_{i}=\arg\min_{1\leq k \leq K}  \|\omg_{1,k}\R_{1,i}-\tilde{\Z}_{1,i}\|_2^2 + \eta_1^2\|\tilde{\Z}_{1,i}\|_0 
\\
+ \|\omg_{2,\hat{l}_{i}}(\omg_{1,k}\R_{1,i}-\tilde{\Z}_{1,i})-\Z_{2,i}\|_2^2
\end{aligned}
%\vspace{-0.05in}
\end{equation}
where $\tilde{\Z}_{1,i} = H_{\eta_1/\sqrt{2}}(\omg_{1,k}\R_{1,i}-0.5\omg_{2,\hat{l}_i}^T\Z_{2,i})$ and $\hat{l}_i$ denotes the \emph{fixed} cluster membership in the second layer. The optimal $\Z_{1,i}$ is then given as follows:
\vspace{-0.15in}
\begin{equation}\label{eq:update_Z1}
\hat{\Z}_{1,i} = H_{\eta_1/\sqrt{2}}(\omg_{1,\hat{k}_i} \R_{1,i} - 0.5 \omg_{2, \hat{l}_i}^T \Z_{2,i}), \quad\forall i.
\vspace{-0.25in}
\end{equation}
%\end{small}
\vspace{-0.20in}
\subsubsection{Update of Transforms in the First Layer}
\vspace{-0.05in}
In this step, we solve subproblem \eqref{eq:sub_P0_omg1} for the transforms $\{\Omega_{1,k}\}$ with the other variables fixed.
\vspace{-0.20in}
\begin{equation}\label{eq:sub_P0_omg1}
%\vspace{-0.05in}
\begin{aligned}
&\min_{\{\omg_{1,k}\}} \sum_{k=1}^{K} \sum_{i \in C_{1,k}} \|\omg_{1,k}\R_{1,i}-\Z_{1,i}\|_2^2 
\\
+ \sum_{k=1}^{K}\sum_{l=1}^{L} &\sum_{j\in C_{2,l}\cap C_{1,k}}  \|\omg_{2,l}(\omg_{1,k}\R_{1,j}-\Z_{1,j})-\Z_{2,j}\|_2^2.
%\quad \mathrm{s.t.} \quad \omg_{1,k}\omg_{1,k}^T=\I, \forall k.
\end{aligned}
%\vspace{-0.05in}
\end{equation}
The problem decouples into $K$ parallel updates, one for each $\omg_{1,k}$.
%The $k$th subproblem is shown as \eqref{eq:sol_P0_omg1}.
%\begin{equation}\label{eq:sol_P0_omg1}
%\begin{aligned}
%&\min_{\omg_{1,k}} \sum_{i \in C_{1,k}} \bigg\{ \|\omg_{1,k}\R_{1,i}-\Z_{1,i}\|_2^2\bigg\}
%\\
%&+ \sum_{l=1}^{l} \sum_{j\in C_{2,l}} \bigg\{ \|\omg_{2,l}\R_{2,j}-\Z_{2,j}\|_2^2\bigg\}
%%\quad \mathrm{s.t.} \quad \omg_{1,k}\omg_{1,k}^T=\I, j \in C_{2,l}.
%\end{aligned}
%\end{equation}
%Let $\R_{1,C_{1,k}}$ and $\Z_{1,C_{1,k}}$ be matrices with columns $\R_{1,i}$ and $\Z_{1,i}$, $i \in C_{1,k}$, respectively, and let $\R_{1,C_{2,l}\cap C_{1,k}}$, $\Z_{1,C_{2,l}\cap C_{1,k}}$, and $\Z_{2,C_{2,l}\cap C_{1,k}}$ be matrices defined similarly.
Let $\R_{1,C_{1,k}}$ and $\Z_{1, C_{1,k}}$, be matrices with columns $\R_{1,i}$, $\Z_{1,i}$, $i \in C_{1,k}$, respectively, and let $\Z_{2,C_{2,l}\cap C_{1,k}}$ be matrix defined similarly. 
%\blue{Operator $\bigcup$ denotes a union for matrices.}  
Then, denoting the full singular value decomposition (SVD) of $\R_{1,C_{1,k}}(\Z_{1,C_{1,k}}^T
+0.5\mathbf{Q}_k^T)$ 
as $\U_{1,k}\Sig_{1,k}\V_{1,k}^T$, where $\mathbf{Q}_k$ is defined in \eqref{eq:Q_k},
%$\R_{1,C_{1,k}}\Z_{1,C_{1,k}}^T+0.5\sum_{l=1}^{L}\R_{1,C_{2,l}\cap C_{1,k}}\Z_{2,C_{2,l}\cap C_{1,k}}^T\omg_{2,l}$ as $\U_{1,k}\Sig_{1,k}\V_{1,k}^T$
%\blue{matrix $\bm{\cup}^{i\in C_{1,k}}\R_{1,i}\Z_{1,i}^T+0.5\R_{1,i}\Z^T_{2,i}\omg_{2,\hat{l}_i}$}
the optimal solution is $\hat{\omg}_{1,k}=\V_{1,k}\U_{1,k}^T$~\cite{ravishankar:15:lst}.
\vspace{-0.1in}
\begin{equation}\label{eq:Q_k}
\begin{aligned}
\mathbf{Q}_k = &[\omg_{2,1}^T\Z_{2,C_{2,1}\cap C_{1,k}},\omg_{2,2}^T\Z_{2,C_{2,2}\cap C_{1,k}},
\\&\cdots,\omg_{2,L}^T\Z_{2,C_{2,L}\cap C_{1,k}}]
\end{aligned}
\vspace{-0.3in}
\end{equation}
\vspace{-0.15in}
%\begin{equation}
%\mathbf{G}_1 = 2\R_{1,C_{1,k}}\Z_{1,C_{1,k}}^T+\sum_{l=1}^{L}\R_{1,C_{2,l}\cap C_{1,k}}\Z_{2,C_{2,l}\cap C_{1,k}}^T\omg_{2,l}
%\end{equation}
%\vspace{-0.45in}
\subsubsection{Update Coefficients and Clusters in the Second Layer}
\vspace{-0.05in}
Next, we solve subproblem \eqref{eq:sub_P0_Z2} with  $\{\Z_{1,i},C_{1,k}$, $\omg_{1,k}$, $\omg_{2,l}\}$ fixed. The (joint) optimal solution for $\{\Z_{2,j},C_{2,l}\}$ can be exactly computed as shown in \eqref{eq:update_C2} and \eqref{eq:update_Z2}.
%shown in \eqref{eq:update_Z2} and \eqref{eq:update_C2}.
\vspace{-0.15in}
\begin{equation}\label{eq:sub_P0_Z2}
\min_{\big\{\substack{\Z_{2,j},\\C_{2,l}}\big\}} \sum_{l=1}^{L} \sum_{j \in C_{2,l}}  \|\omg_{2,l}\R_{2,j}-\Z_{2,j}\|_2^2 + \eta_2^2\|\Z_{2,j}\|_0 .
\vspace{-0.05in}
\end{equation}
%\begin{equation}\label{eq:update_Z2}
%\hat{\Z}_{2,j}=H_{\eta_2}(\omg_{2,l}\R_{2,j}), \quad \forall j \in C_{2,l}.
%\vspace{-0.05in}
%\end{equation}
\vspace{-0.15in}
\begin{equation}\label{eq:update_C2}
\hat{l}_{j}\hspace{-0.05in}=\hspace{-0.03in}\arg\hspace{-0.03in}\min_{1\leq l \leq L} \hspace{-0.03in} \|\omg_{2,l}\R_{2,j}-\tilde{\Z}_{2,j}\|_2^2 + \eta_2^2\|\tilde{\Z}_{2,j}\|_0 , \; \forall j \in C_{2,l},
\vspace{-0.05in}
\end{equation}
%\vspace{-0.05in}
where $\tilde{\Z}_{2,j} = H_{\eta_2}(\omg_{2,l}\R_{2,j})$. Then the optimal solution to $\Z_{2,j}$ is as follows:
\vspace{-0.15in}
\begin{equation}\label{eq:update_Z2}
\hat{\Z}_{2,j} = H_{\eta_2}(\omg_{2,\hat{l}_j} \R_{2,j}), \quad\forall j.
\vspace{-0.20in}
\end{equation}
\vspace{-0.25in}
\subsubsection{Update of Transforms in the Second Layer}
\vspace{-0.05in}
In this step, we solve the following subproblem for {$\omg_{2,l}$}, with the other variables kept fixed:
\vspace{-0.10in}
\begin{equation}\label{eq:sub_P0_omg2}
\min_{\{\omg_{2,l}\}} \sum_{l=1}^{L} \sum_{j\in C_{2,l}}  \|\omg_{2,l}\R_{2,j}-\Z_{2,j}\|_2^2
\vspace{-0.05in}
%\quad \mathrm{s.t.} \quad \omg_{2,l}\omg_{2,l}^T=\I, \forall l.
\end{equation}
%Since above problem could be separated into $l$ independent single transform learning problems, we could solve in parallel. The $l$th such optimization problem is as follows:
%\begin{equation}\label{eq:sol_P0_omg2}
%\min_{\omg_{2,l}} \sum_{j\in C_{2,l}} \bigg\{ \|\omg_{2,j}\R_{2,j}-\Z_{2,j}\|_2^2\bigg\}
%\quad \mathrm{s.t.} \quad \omg_{2,l}\omg_{2,l}^T=\I.
%\end{equation}
Problem \eqref{eq:sub_P0_omg2} decouples into $L$ different updates, one for each transform. Let $\R_{2,C_{2,l}}$ and $\Z_{2,C_{2,l}}$ be matrices with columns $\R_{2,j}$ and $\Z_{2,j}$, $j \in C_{2,l}$, respectively. Then, denoting the full SVD of $\R_{2,C_{2,l}}\Z_{2,C_{2,l}}^T$ as $\U_{2,l}\Sig_{2,l}\V_{2,l}^T$, the optimal $\hat{\omg}_{2,l}=\V_{2,l}\U_{2,l}^T$.
\vspace{-0.15in}
\section{Approach for Image Reconstruction}
\vspace{-0.10in}
\subsection{CT Reconstruction Formulation}
\vspace{-0.05in}
After learning the collections of transforms $\{\omg_{1,k}\}$ and $\{\omg_{2,l}\}$, the learned MCST2 model is incorporated into the reconstruction problem via a data-driven regularizer. We then reconstruct the (vectorized) image $\x \in \mathbb{R}^{N_p}$ from noisy sinogram measurements $\y \in \mathbb{R}^{N_d}$ by solving the following problem:
\vspace{-0.10in}
\begin{equation}\label{eq:P1}	
\min_{\x \succeq \0}  \frac{1}{2}\|\y - \A \x\|^2_{\W} + \beta\mathrm{S}(\x),
\tag{P1}
\vspace{-0.07in}
\end{equation}
where the regularizer $\mathrm{S}(\x)$ is defined as follows, with $\mathcal{H}_1 = \{\Z_{1,i},\Z_{2,j},C_{1,k},C_{2,l}\}$ again denoting the set of \emph{all} optimized variables:
\vspace{-0.10in}
\begin{equation}\label{eq:Rx_UMRST}
\vspace{-0.07in}
\begin{aligned}
%&\mathrm{R}(\x) \triangleq  
&\min_{\mathcal{H}_1}  \sum_{k=1}^{K} \sum_{i\in C_{1,k}}  \|\omg_{1,k}\R_{1,i}\hspace{-0.02in}-\hspace{-0.02in}\Z_{1,i}\|_2^2 \hspace{-0.02in}+\hspace{-0.02in} \gamma_1^2\|\Z_{1,i}\|_0 
\\
&  \quad  +  \sum_{l=1}^{L} \sum_{j\in C_{2,l}}  \|\omg_{2,l}\R_{2,j}-\Z_{2,j}\|_2^2 + \gamma_2^2\|\Z_{2,j}\|_0 ,
\\
& \mathrm{s.t.} \; \R_{1,i}=\P_{i}\x, \;\R_{2,i}= \omg_{1,k}\R_{1,i}-\Z_{1,i},\; \forall i \in C_{1,k}, \forall k.
%\\
%&\quad \quad \quad \{C_{1,k}\} \in \mathcal{G}_1, \{C_{2,l}\} \in \mathcal{G}_2.
\end{aligned}
%\vspace{-0.05in}
\end{equation}	
In problem \eqref{eq:P1}, $\A \in \mathbb{R}^{N_d\times N_p}$ denotes the CT measurement matrix, and $\W \in \mathbb{R}^{N_d\times N_d}$ is a diagonal weighting matrix, whose diagonal elements are the estimated inverse variances of elements of $\y$. The operator $\P_{i}$ extracts the $i$th vectorized patch of $\x$ as $\P_{i}\x$. The parameter $\beta$ denotes the regularizer weighting, and $\gamma_1$ and $\gamma_2$ are non-negative parameters that control the sparsity levels of the sparse coefficients.
\vspace{-0.15in}
\subsection{Image Reconstruction Algorithm}
\vspace{-0.10in}
Similar to the learning algorithm, we use an exact block coordinate descent (BCD) algorithm to optimize \eqref{eq:P1}. The algorithm cycles over updates of the image $\x$, sparse coefficients $\Z_1$ and $\Z_2$, and cluster memberships $\{C_{1,k}\}$, $\{C_{2,l}\}$. The algorithm enforces monotone decrease of the underlying objective.
\vspace{-0.15in}
\subsubsection{Image Update Step}
\vspace{-0.10in}
In this step, we update $\x$ in \eqref{eq:P1} with the other variables fixed, which leads to the subproblem \eqref{eq:image_update}. We use the efficient relaxed LALM (rLALM) algorithm~\cite{nien:16:rla} to solve \eqref{eq:image_update}. The detailed description of this algorithm can be found in \cite{yang:20:mars}. 
\vspace{-0.10in}
\begin{equation}
\label{eq:image_update} 
\min_{\x \succeq \0} \frac{1}{2} \|\y - \A\x \|^2_{\W} + \beta\mathrm{S}_1(\x), 
\vspace{-0.05in}
\end{equation}
where $\mathrm{S}_1(\x) \triangleq  \sum_{k=1}^{K} \sum_{i\in C_{1,k}}  \|\omg_{1,k}\P_{i}\x-\Z_{1,i}\|_2^2  +  \sum_{l=1}^{L} \sum_{j\in C_{2,l}}  \|\omg_{2,l}(\omg_{1,k}\P_{j}\x-\Z_{1,j})-\Z_{2,j}\|_2^2 $.
%The rLALM algorithm requires the matrix $\D_{\mathrm{S}_1}$, which is the diagonal Hessian matrix of the regularizer $\mathrm{S}_1(\x)$, and is precomputed as follows to save runtime. 
%\begin{equation}\label{eq:D_R}	
%\begin{aligned}		
%\D_{\mathrm{S}_1}  \triangleq 2 \sum_{k=1}^{K} \sum_{i\in C_{1,k}} \hspace{-0.10in} \P_{i}^T\P_{i} 
%+ 2 \sum_{l=1}^{L} \sum_{j\in C_{2,l}} \hspace{-0.10in} \P_{j}^T\P_{j} 
%\end{aligned}
%\vspace{-0.05in}
%\end{equation}	
%If overlapping image patches are selected with a patch stride of $1$ pixel along each direction, and including patches that wrap around image boundaries~\cite{zheng:18:pua}, then $\D_{\mathrm{S}_1} = 4\P\I$, where $\I$ is the identity matrix.

\vspace{-0.10in}
\subsubsection{Sparse Coding and Clustering Step}
\vspace{-0.05in}
With $\x$ fixed, \eqref{eq:P1} is reduced to the same subproblems as \eqref{eq:sub_P0_Z1} and \eqref{eq:sub_P0_Z2}. Then $\{\Z_{1,i}, C_{1,k}\}$ and $\{\Z_{2,j}, C_{2,l}\}$ are updated in the same manner as in \eqref{eq:update_C1}, \eqref{eq:update_Z1},  \eqref{eq:update_C2}, and \eqref{eq:update_Z2}.
\vspace{-0.15in}
\section{Experiments}
\vspace{-0.10in}
\subsection{Experiment Setup}
\vspace{-0.10in}
We study the performance of MCST2 for the XCAT phantom and Mayo Clinic data. For XCAT phantom case, the low-dose measurements are simulated from the groundtruth image with GE 2D  LightSpeed fan-beam geometry corresponding to a monoenergetic source. 
For Mayo Clinic data case, we simulated the low-dose measurements from the regular-dose images with a fan-beam CT geometry corresponding to a monoenergetic source. 
The width of each detector column is $1.2858$ mm, the distances from source to detector, source to rotation center are $1085.6$ mm and $595$ mm, respectively.
We set the incident photon intensity $I_0=1\times 10^4$  per ray and with no scatter. The ``Possion + Gaussian'' noisy model~\cite{ding:16:mmp} is used to generate synthesized low-dose measurements of size $888\times 984$ for the XCAT phantom and $736\times 1152$ for Mayo Clinic data, respectively. 
Two types of metrics (RMSE and SSIM) are applied for evaluating image reconstruction quality. We compute the root mean square error (RMSE) and the structural similarity index measure (SSIM) in a circular central region of the images, which includes all the tissues. 
%For each reconstruction slice, we achieves the lowest RMSE (HU) and the highest SSIM value.
\vspace{-0.15in}
\subsection{Transform Learning}
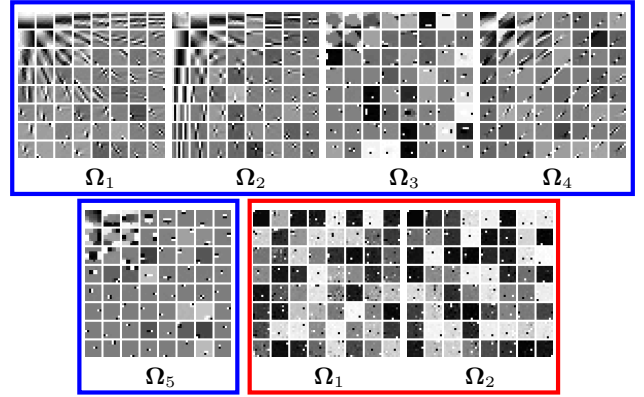
\begin{figure}[!h] 
	\centering  
	\begin{tikzpicture}
	[spy using outlines={rectangle,green,magnification=2,size=9mm, connect spies}]
	\node {
	\includegraphics[width=0.11\textwidth]{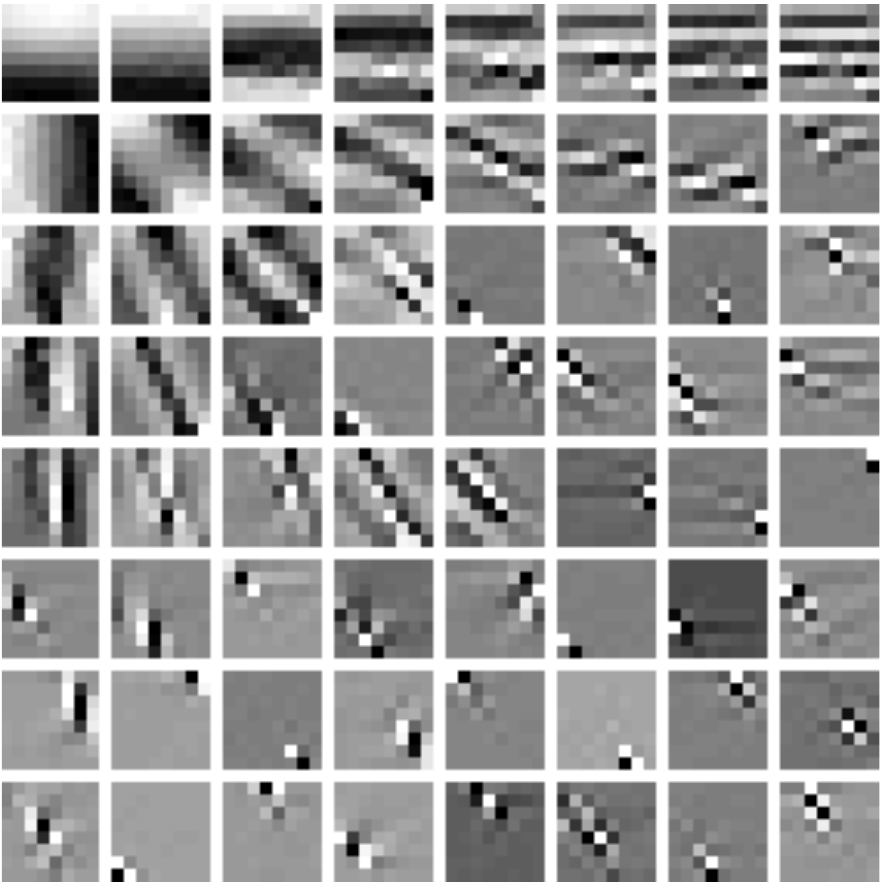}
	\includegraphics[width=0.11\textwidth]{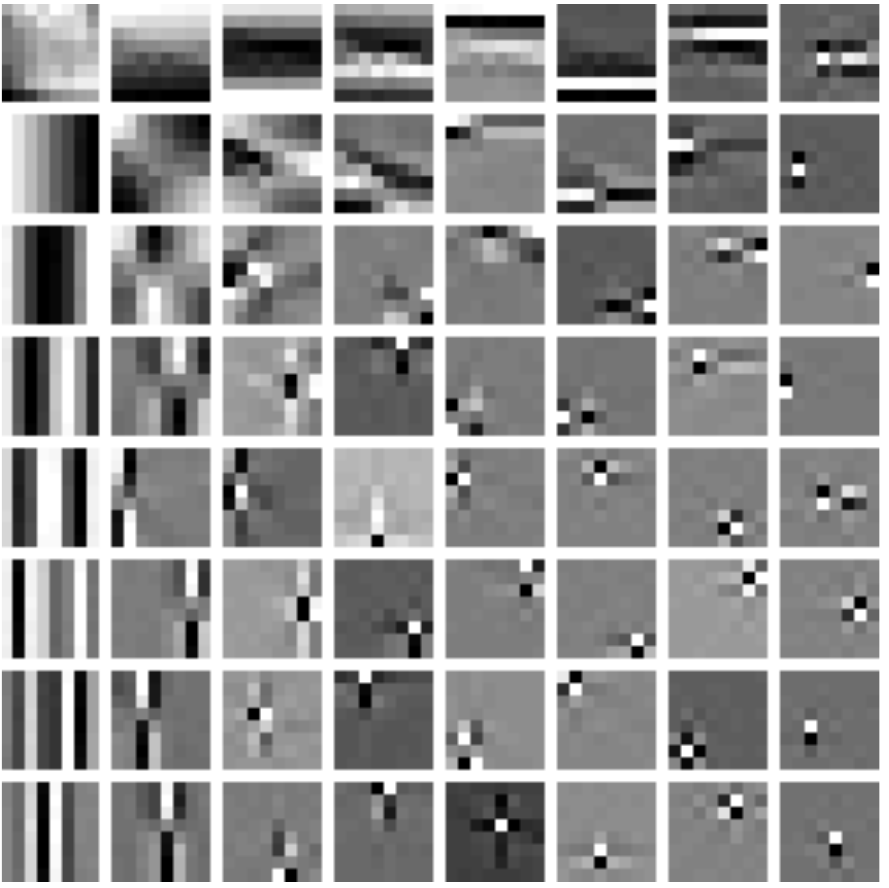}
	\includegraphics[width=0.11\textwidth]{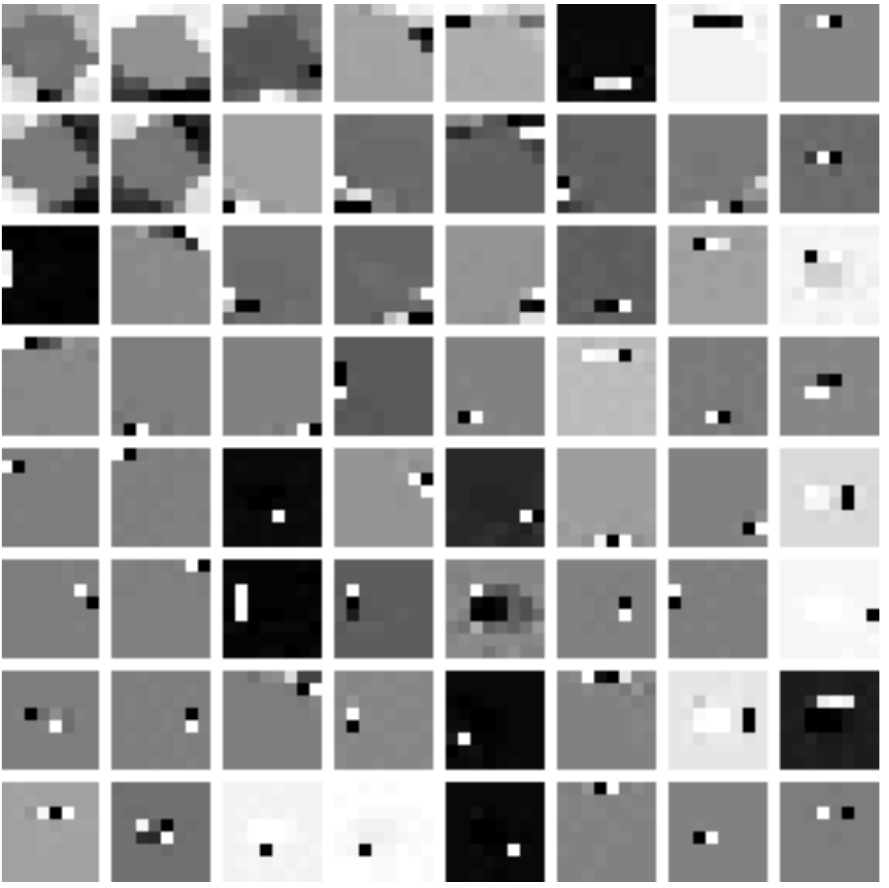}
	\includegraphics[width=0.11\textwidth]{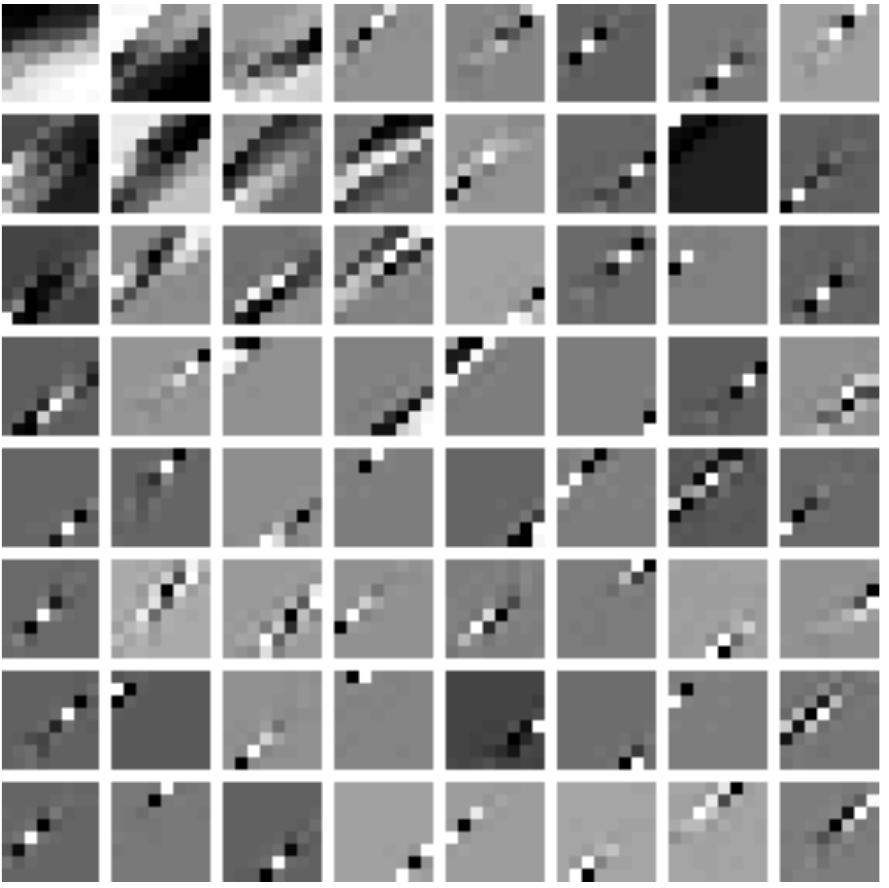}
	};
	\draw[blue, ultra thick] (-4.13,-1.45) rectangle (4.12,1.1);
	\put(-90,-38){ \color{black}{\bf \small{$\omg_1$}}}
	\put(-33,-38){ \color{black}{\bf \small{$\omg_2$}}}
	\put(25,-38){ \color{black}{\bf \small{$\omg_3$}}}
	\put(83,-38){ \color{black}{\bf \small{$\omg_4$}}}
	\end{tikzpicture}
	\\
%	\vspace{0.10in}
	\begin{tikzpicture}
	[spy using outlines={rectangle,green,magnification=2,size=9mm, connect spies}]
	\node {
	\includegraphics[width=0.11\textwidth]{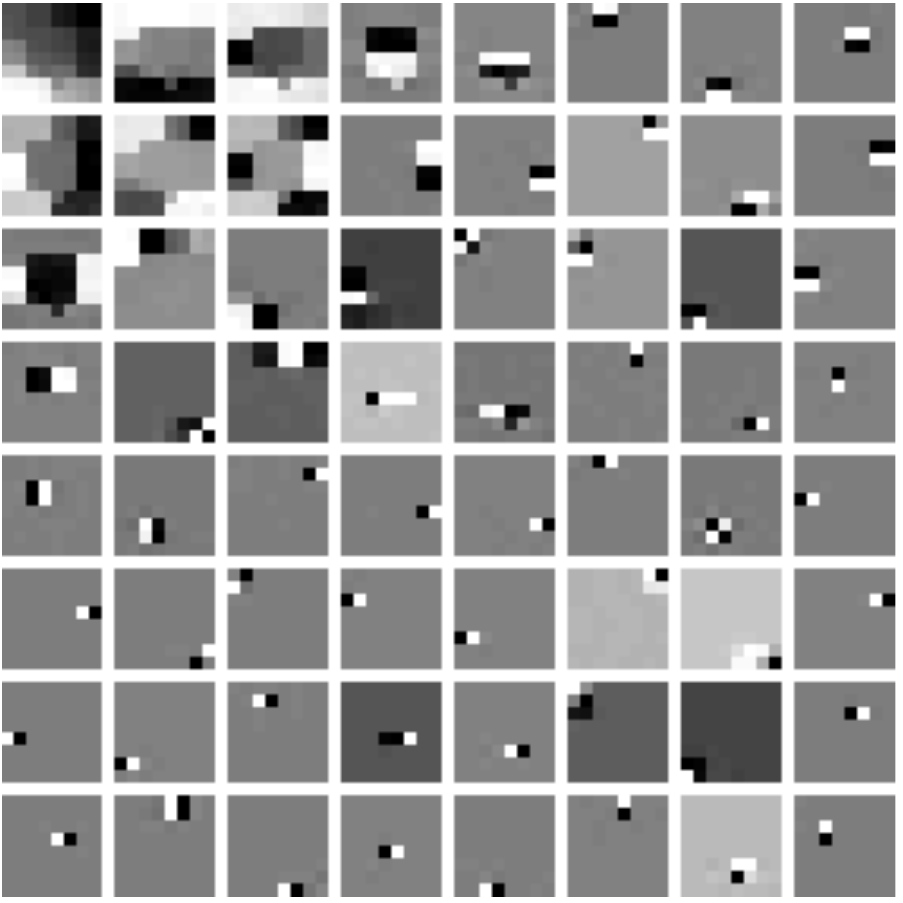}};
	\draw[blue, ultra thick] (-1.05,-1.45) rectangle (1.05,1.1);
	\put(-5,-38){ \color{black}{\bf \small{$\omg_5$}}}
	\end{tikzpicture}
	\hspace{-0.05in}
	\begin{tikzpicture}
	[spy using outlines={rectangle,green,magnification=2,size=9mm, connect spies}]
	\node {
		\includegraphics[width=0.11\textwidth]{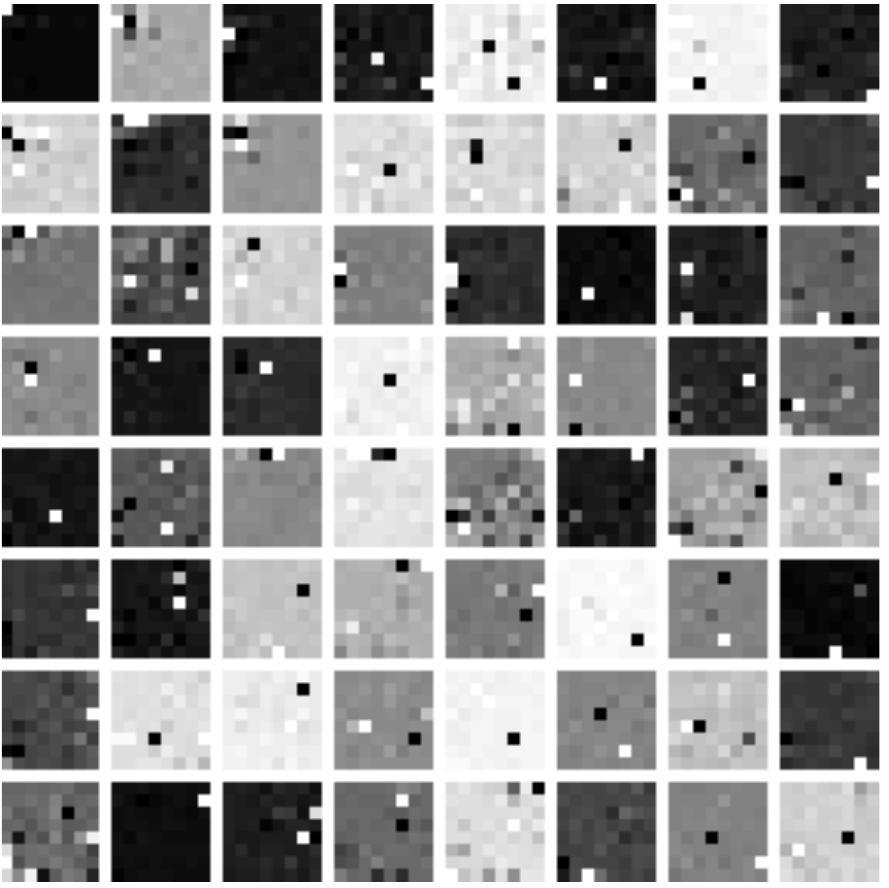}
		\includegraphics[width=0.11\textwidth]{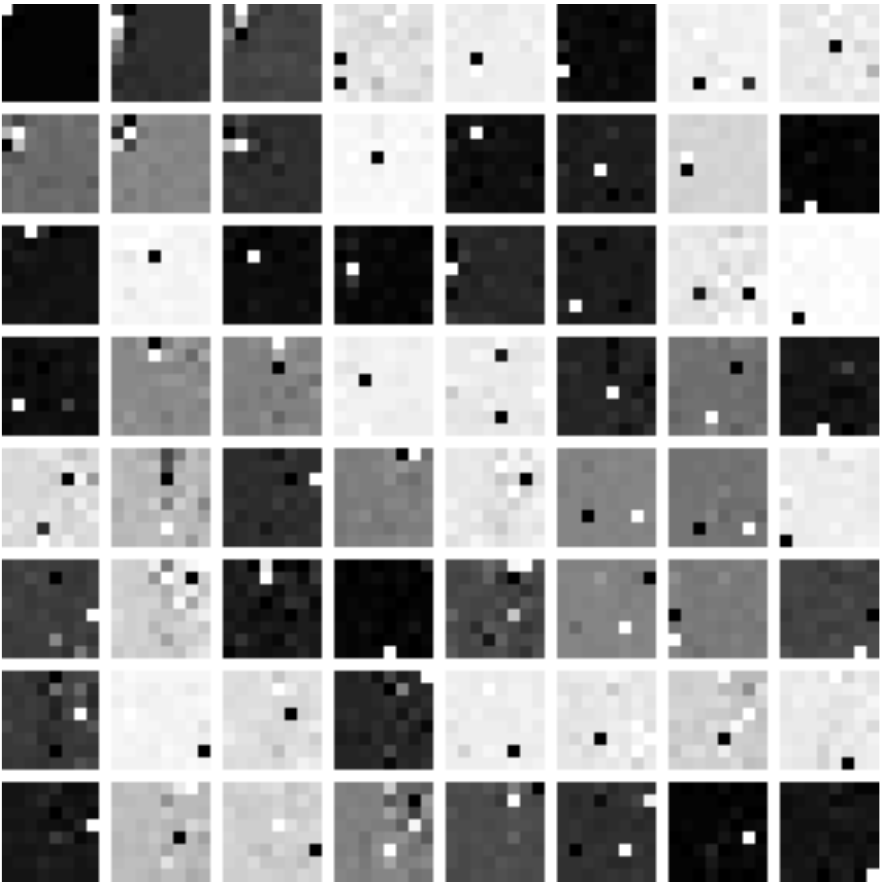}};
	\draw[red, ultra thick] (-2.05,-1.45) rectangle (2.05,1.1);
	\put(-34,-38){ \color{black}{\bf \small{$\omg_1$}}}
	\put(23,-38){ \color{black}{\bf \small{$\omg_2$}}}
	\end{tikzpicture}
	\vspace{-0.10in}
	\caption{Pre-learned transforms from XCAT phantom for the MCST2 model with $\mathbf{5}$ clusters in the first layer (shown in the blue box) and $\mathbf{2}$ clusters in the second layer (shown in the red box). Each row of the transform matrices is displayed as a square $8\times 8$ patch.  }
	\label{fig:tran_XCAT}
	\vspace{-0.15in}
\end{figure}

\vspace{-0.10in}
For the learning stage, we used five $420\times 420$ XCAT phantom slices to train the MCST2 model. We also used seven slices of size $512\times 512$ from the Mayo Clinic data set to learn transforms. We ran $1000$ iterations of the BCD algorithm to ensure convergence. The number of clusters in the two layers were $5$ and $2$, respectively. We set $(\eta_1,\eta_2)$ $=$ $(125,70)$ and $(60,10)$ for the XCAT phantom and Mayo Clinic data, respectively. Fig.~\ref{fig:tran_XCAT} shows the transforms in the MCST2 model that were learned from the XCAT phantom data. Each row of the transform matrices is displayed as an $8\times 8$ square patch. The pre-learned transforms in the first layer (blue box) show oriented and gradient-like features that sparsify the training image patches. For the second layer, the pre-learned transforms (red box) capture finer level features that further sparsify the filtering residuals. 
\vspace{-0.15in}
\vspace{-0.15in}
\subsection{Reconstruction Results}
\begin{figure}[!h]
	\vspace{-0.15in}
	\centering  
	\begin{tikzpicture}
	[spy using outlines={rectangle,green,magnification=2,size=9mm, connect spies}]
	\node {\includegraphics[width=0.23\textwidth]{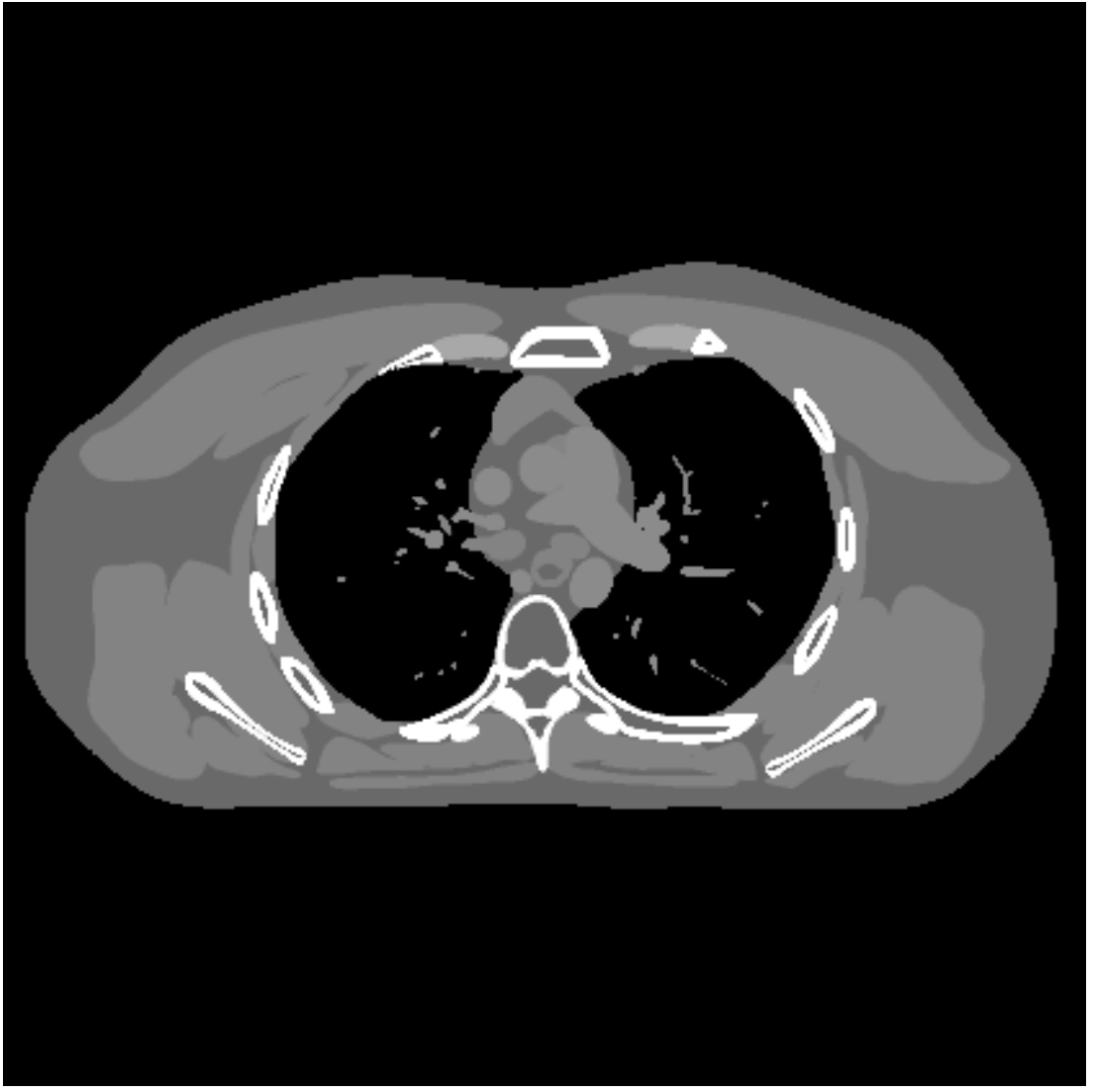}	};
	\spy [green, draw, height = 0.9cm, width = 0.9cm, magnification = 1.5,
	connect spies] on (-0.05,0.20) in node [left] at (2.00,-1.60);
	\spy [green, draw, height = 0.9cm, width = 0.9cm, magnification = 2,
	connect spies] on (-0.05,-0.55) in node [left] at (-1.12,-1.60);	
	%\draw[red] (-1.95,0)--(1.95,0);
	%\draw[blue] (0,-1.95)--(0,1.95);
	\end{tikzpicture}
	\put(-89,110){ \color{white}{\bf \small{RMSE:0.00}}}
	\put(-89,100){ \color{white}{\bf \small{SSIM:1.000}}}
	\put(-87,10){ \color{white}{\bf \small{Reference}}} 
	\hspace{-0.12in}
	\begin{tikzpicture}
	[spy using outlines={rectangle,green,magnification=2,size=9mm, connect spies}]
	\node {\includegraphics[width=0.23\textwidth]{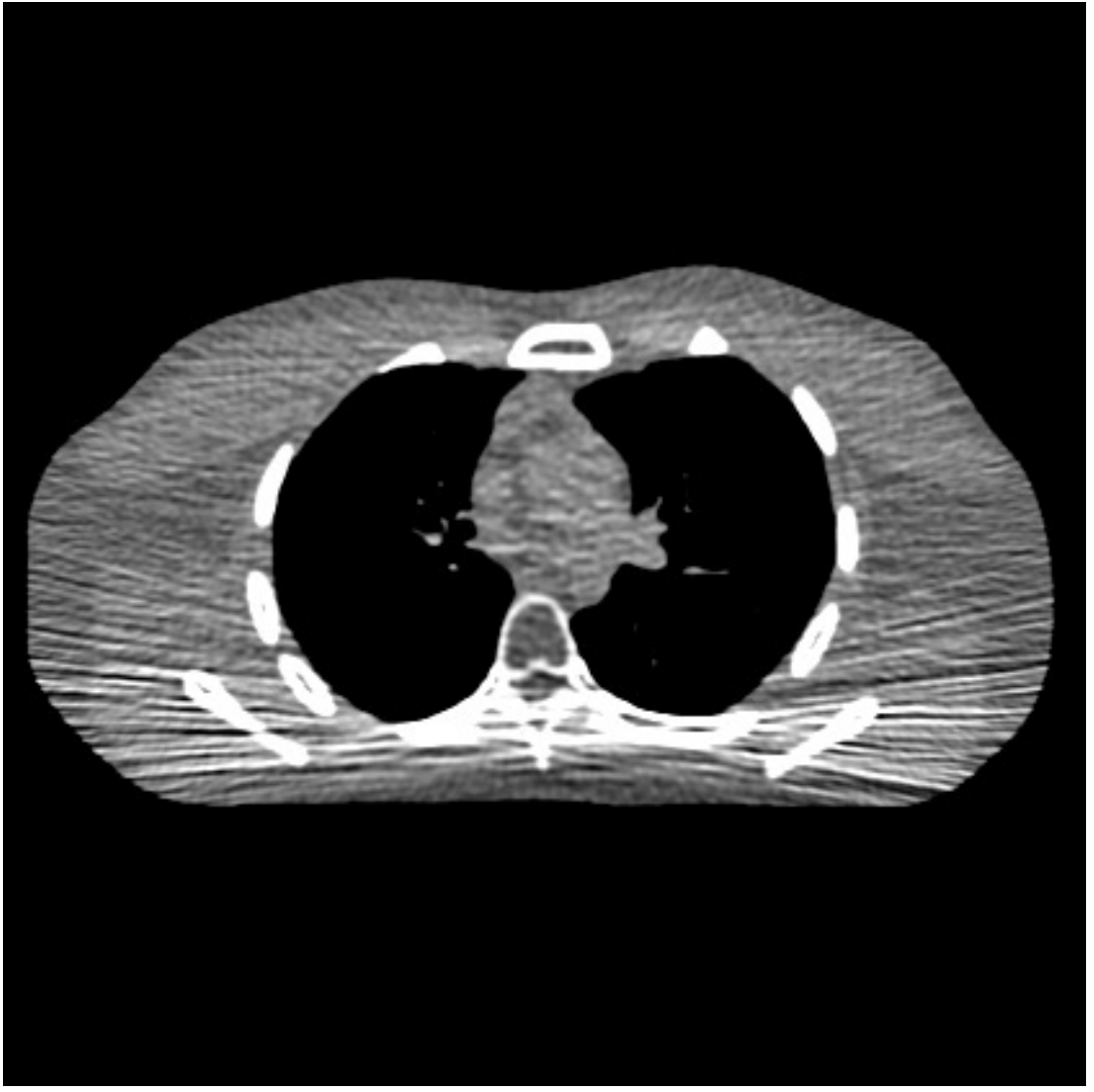}	};
	\spy [green, draw, height = 0.9cm, width = 0.9cm, magnification = 1.5,
	connect spies] on (-0.05,0.20) in node [left] at (2.00,-1.60);
	\spy [green, draw, height = 0.9cm, width = 0.9cm, magnification = 2,
	connect spies] on (-0.05,-0.55) in node [left] at (-1.12,-1.60);	
	\end{tikzpicture}
	\put(-89,110){ \color{white}{\bf \small{RMSE:72.0}}}
	\put(-89,100){ \color{white}{\bf \small{SSIM:0.552}}}
	\put(-75,10){ \color{white}{\bf \small{FBP}}} 
	\\
	\vspace{-0.10in}
	\begin{tikzpicture}
	[spy using outlines={rectangle,green,magnification=2,size=9mm, connect spies}]
	\node {\includegraphics[width=0.23\textwidth]{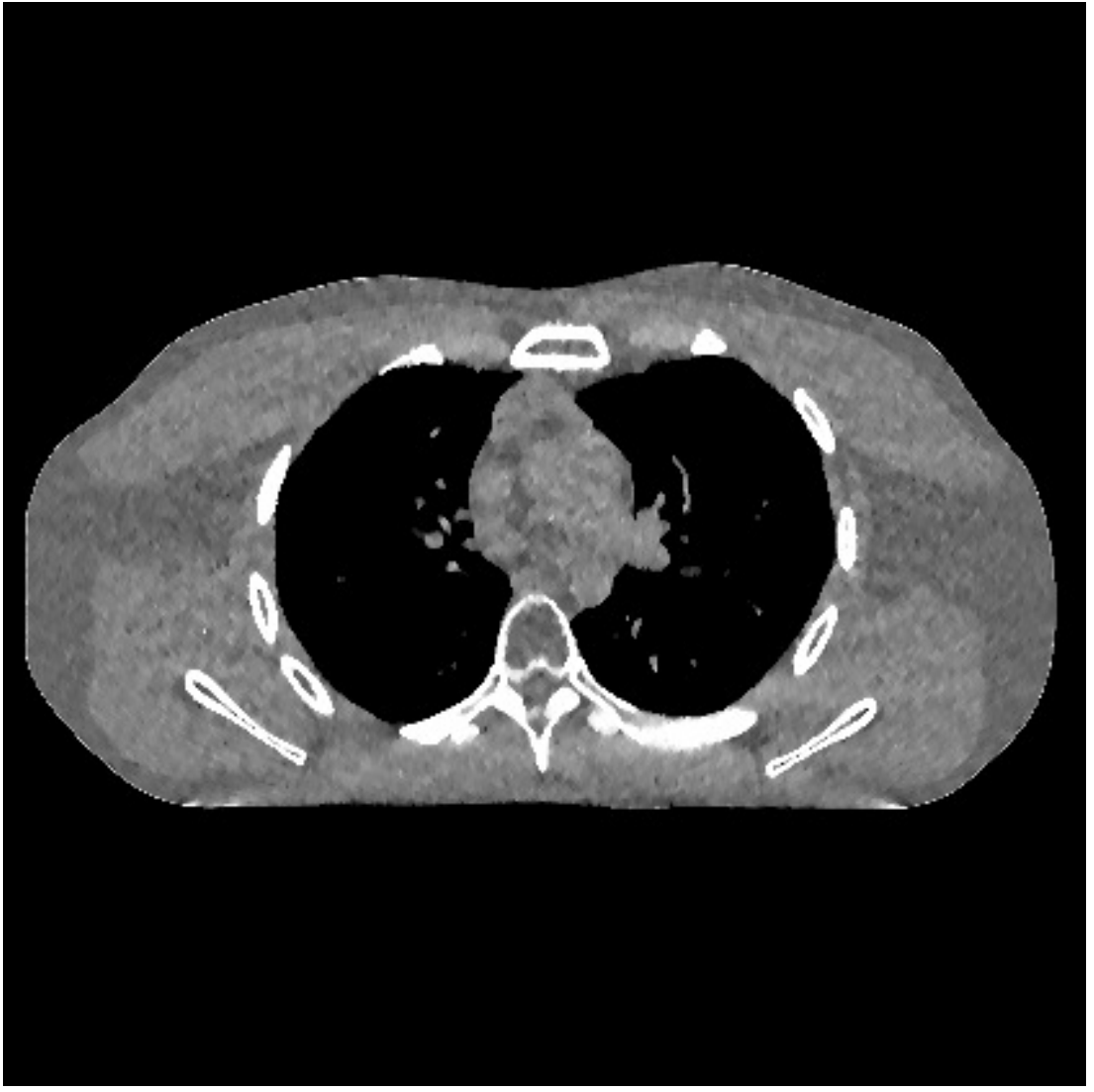}	};
	\spy [green, draw, height = 0.9cm, width = 0.9cm, magnification = 1.5,
	connect spies] on (-0.05,0.20) in node [left] at (2.00,-1.60);
	\spy [green, draw, height = 0.9cm, width = 0.9cm, magnification = 2,
	connect spies] on (-0.05,-0.55) in node [left] at (-1.12,-1.60);	
	\end{tikzpicture}
	\put(-89,110){ \color{white}{\bf \small{RMSE:39.2}}}
	\put(-89,100){ \color{white}{\bf \small{SSIM:0.892}}}
	\put(-70,10){ \color{white}{\bf \small{EP}}}
	\hspace{-0.12in}
	\begin{tikzpicture}
	[spy using outlines={rectangle,green,magnification=2,size=9mm, connect spies}]
	\node {\includegraphics[width=0.23\textwidth]{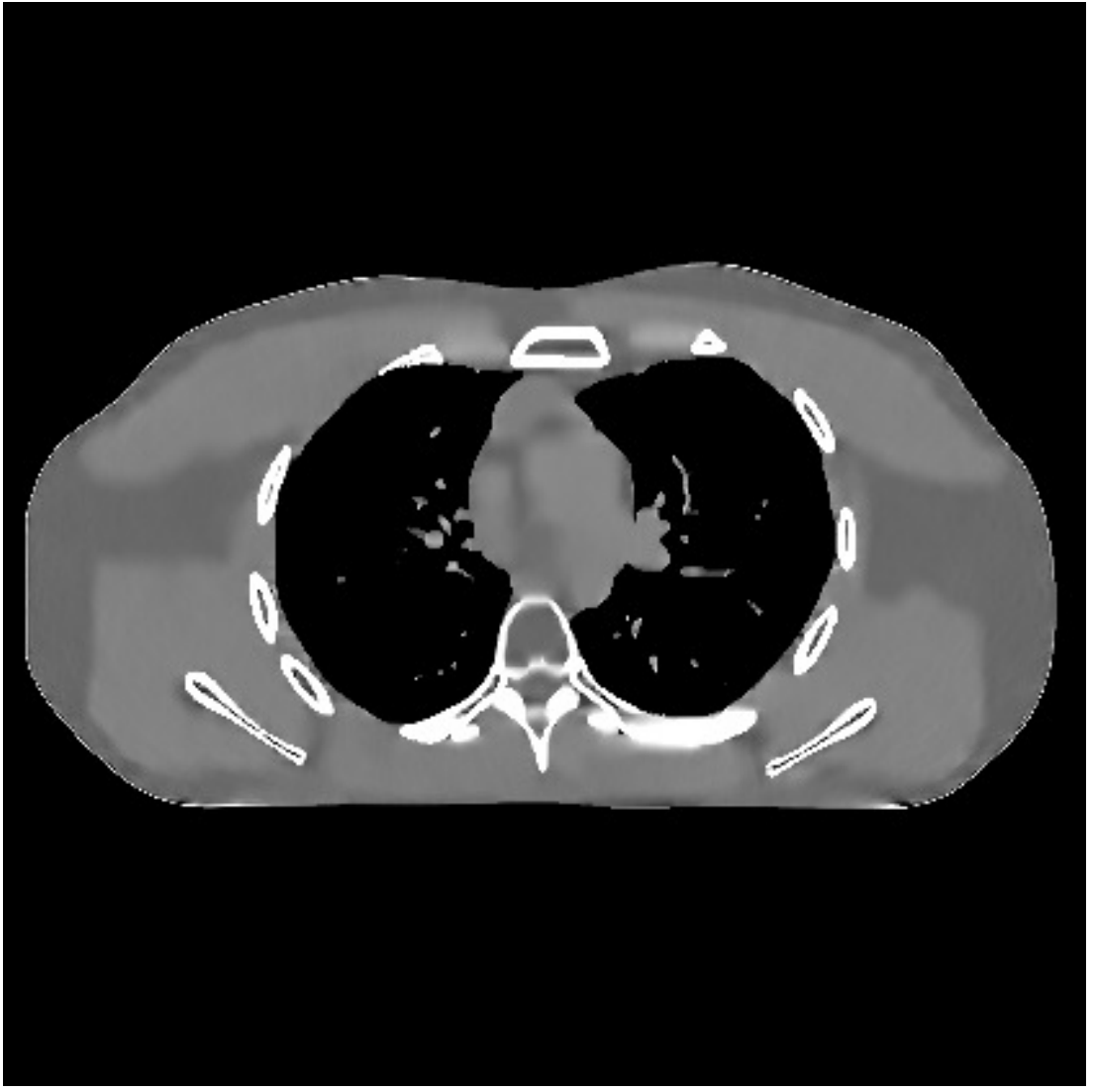}	};
	\spy [green, draw, height = 0.9cm, width = 0.9cm, magnification = 1.5,
	connect spies] on (-0.05,0.20) in node [left] at (2.00,-1.60);
	\spy [green, draw, height = 0.9cm, width = 0.9cm, magnification = 2,
	connect spies] on (-0.05,-0.55) in node [left] at (-1.12,-1.60);	
	\end{tikzpicture}
	\put(-89,110){ \color{white}{\bf \small{RMSE:34.6}}}
	\put(-89,100){ \color{white}{\bf \small{SSIM:0.970}}}
	\put(-82,10){ \color{white}{\bf \small{MRST2}}}
	\\
	\vspace{-0.10in}
	\begin{tikzpicture}
	[spy using outlines={rectangle,green,magnification=2,size=9mm, connect spies}]
	\node {\includegraphics[width=0.23\textwidth]{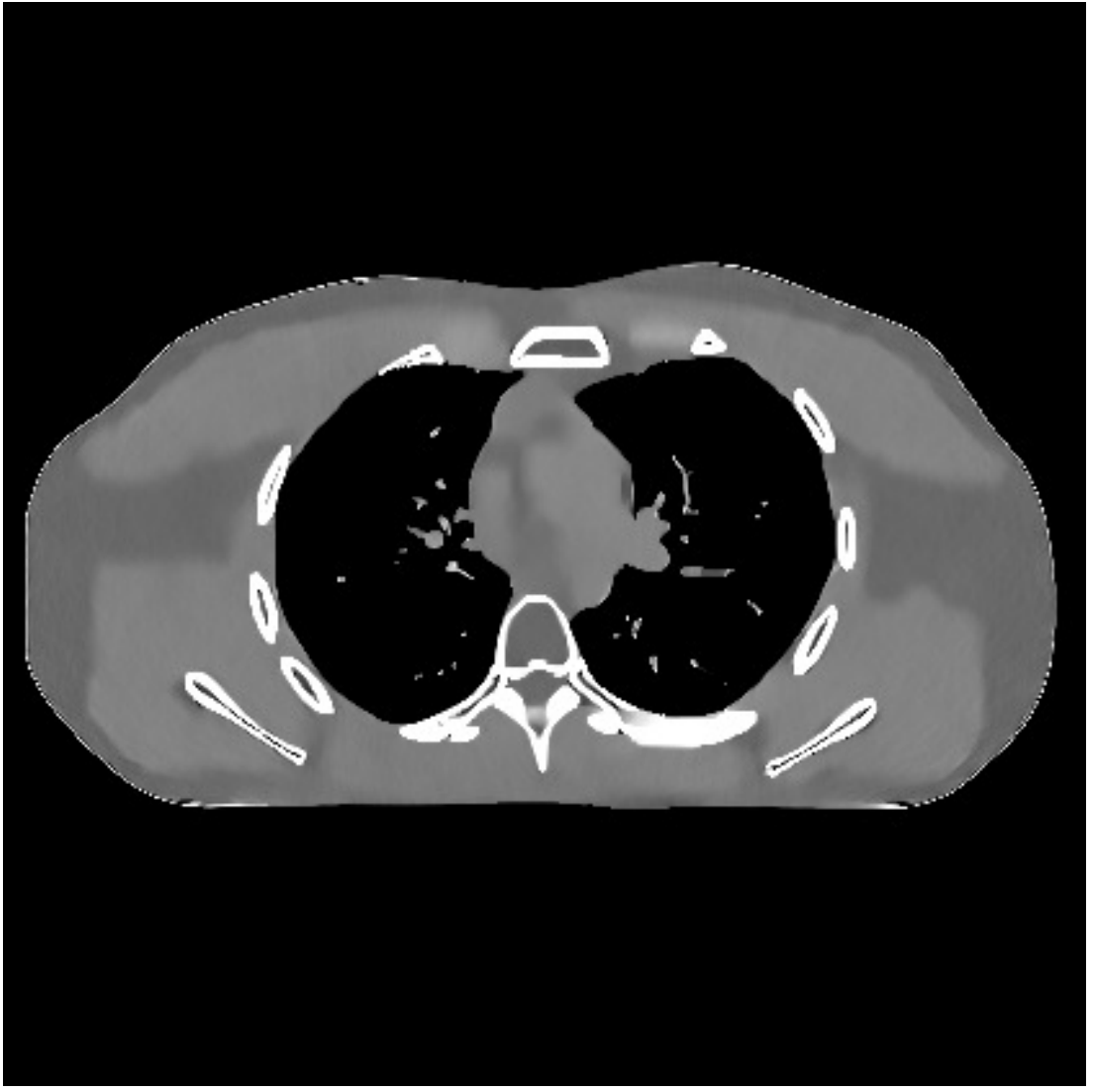}	};
	\spy [green, draw, height = 0.9cm, width = 0.9cm, magnification = 1.5,
	connect spies] on (-0.05,0.20) in node [left] at (2.00,-1.60);
	\spy [green, draw, height = 0.9cm, width = 0.9cm, magnification = 2,
	connect spies] on (-0.05,-0.55) in node [left] at (-1.12,-1.60);	
	\end{tikzpicture}
	\put(-89,110){ \color{white}{\bf \small{RMSE:34.0}}}
	\put(-89,100){ \color{white}{\bf \small{SSIM:0.968}}}
	\put(-82,10){ \color{white}{\bf \small{ULTRA}}}
	\hspace{-0.12in}
	\begin{tikzpicture}
	[spy using outlines={rectangle,green,magnification=2,size=9mm, connect spies}]
	\node {\includegraphics[width=0.23\textwidth]{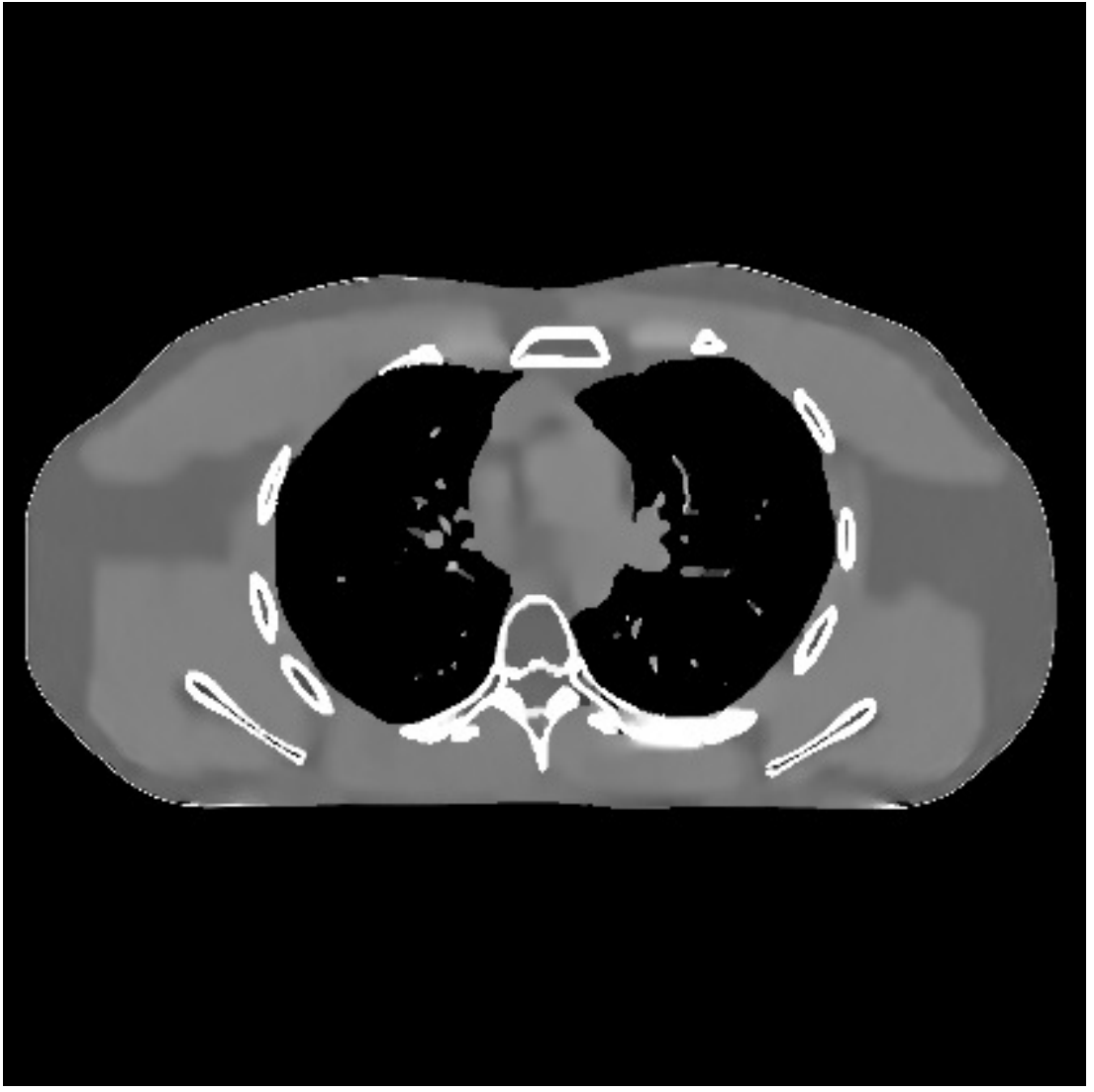}	};
	\spy [green, draw, height = 0.9cm, width = 0.9cm, magnification = 1.5,
	connect spies] on (-0.05,0.20) in node [left] at (2.00,-1.60);
	\spy [green, draw, height = 0.9cm, width = 0.9cm, magnification = 2,
	connect spies] on (-0.05,-0.55) in node [left] at (-1.12,-1.60);	
	\end{tikzpicture}
	\put(-89,110){ \color{green}{\bf \small{RMSE:33.5}}}
	\put(-89,100){ \color{green}{\bf \small{SSIM:0.973}}}
	\put(-82,10){ \color{white}{\bf \small{MCST2}}}
	\vspace{-0.15in}
	\caption{Comparison of reconstructions of one slice of the XCAT phantom with the FBP, PWLS-EP, PWLS-MRST2, PWLS-ULTRA, and PWLS-MCST2 methods, respectively, at incident photon intensity $I_0=1\times 10^{4}$. The display window is [800, 1200] HU.
	}
	\vspace{-0.20in}
	\label{fig:recon_XCAT_slice48}
\end{figure}
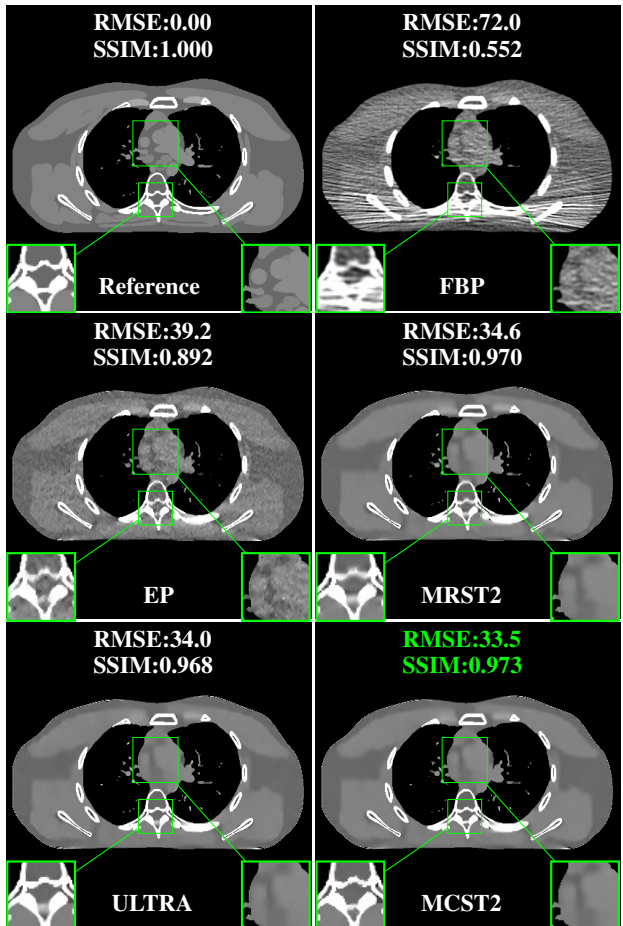

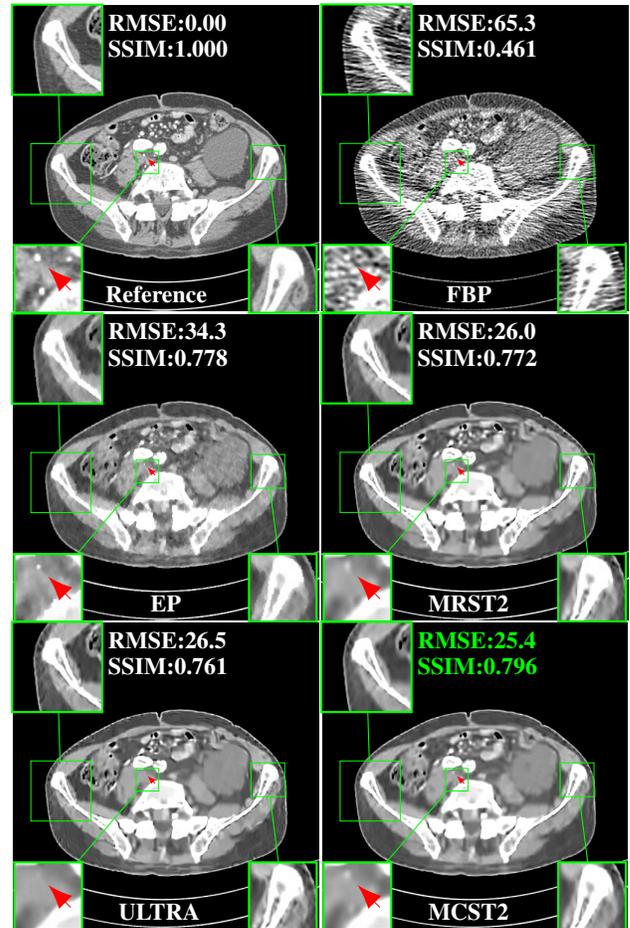
\begin{figure}[!h]
	\vspace{-0.1in}
	\centering  
	\begin{tikzpicture}
	[spy using outlines={rectangle,green,magnification=2,size=9mm, connect spies}]
	\node {\includegraphics[width=0.23\textwidth]{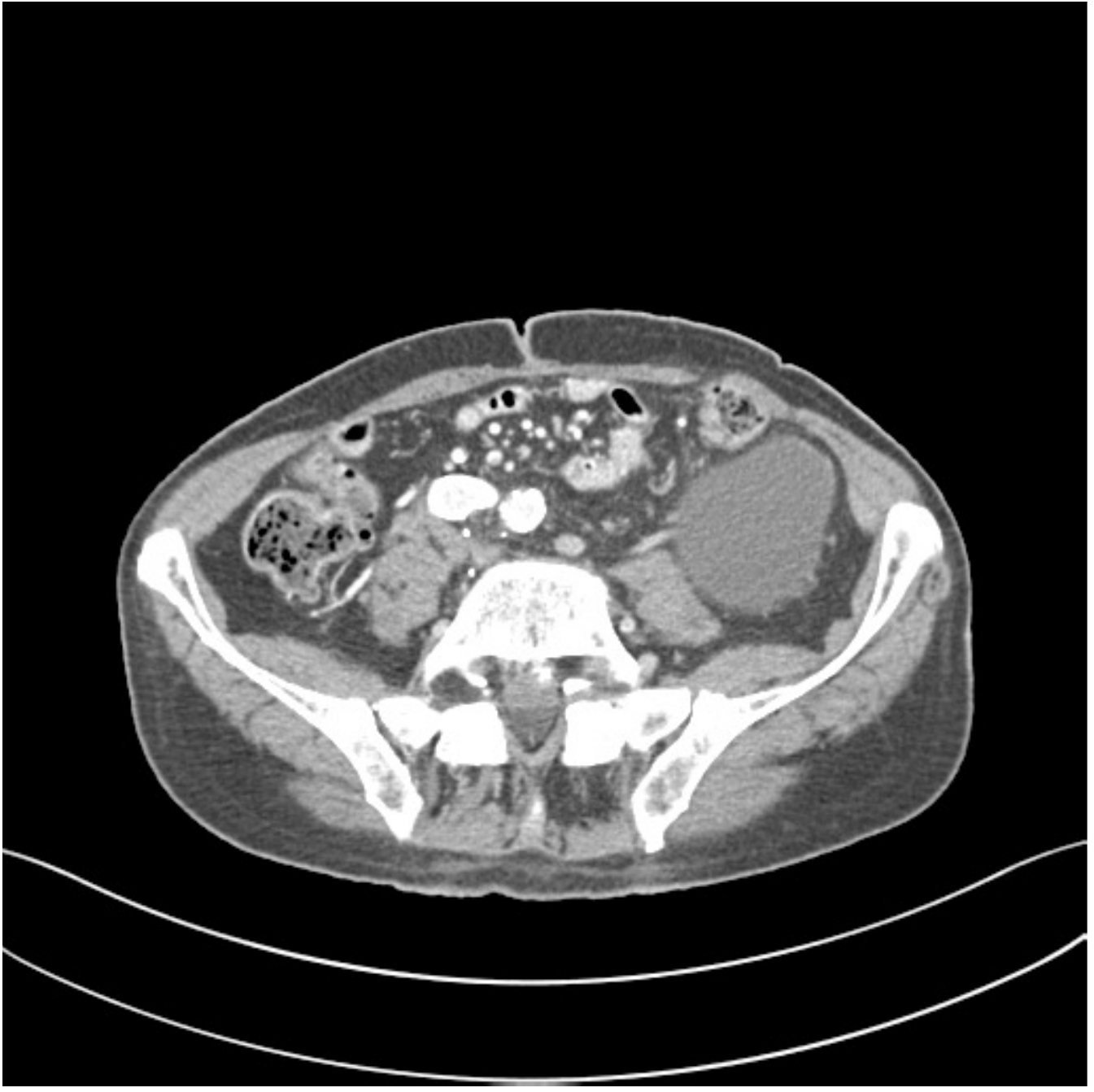}	};
	\spy [green, draw, height = 0.9cm, width = 0.9cm, magnification = 2,
	connect spies] on (1.35,-0.05) in node [left] at (2.00,-1.60);
	\spy [green, draw, height = 0.9cm, width = 0.9cm, magnification = 3,
	connect spies] on (-0.25,-0.05) in node [left] at (-1.12,-1.60);
	\spy [green, draw, height = 1.2cm, width = 1.2cm, magnification = 1.5,
	connect spies] on (-1.40,-0.20) in node [left] at (-0.85,1.45);	
	\draw[line width=1pt, ultra thin, -latex, red] (-0.15,-0.10) -- node[xshift=-0.05cm,yshift=-0.09cm] {\tiny{}} (-0.25,-0.00);
	\end{tikzpicture}
	\put(-86,110){ \color{white}{\bf \small{RMSE:0.00}}}
	\put(-86,100){ \color{white}{\bf \small{SSIM:1.000}}}
	\put(-87,7){ \color{white}{\bf \small{Reference}}} 
	\hspace{-0.12in}
	\begin{tikzpicture}
	[spy using outlines={rectangle,green,magnification=2,size=9mm, connect spies}]
	\node {\includegraphics[width=0.23\textwidth]{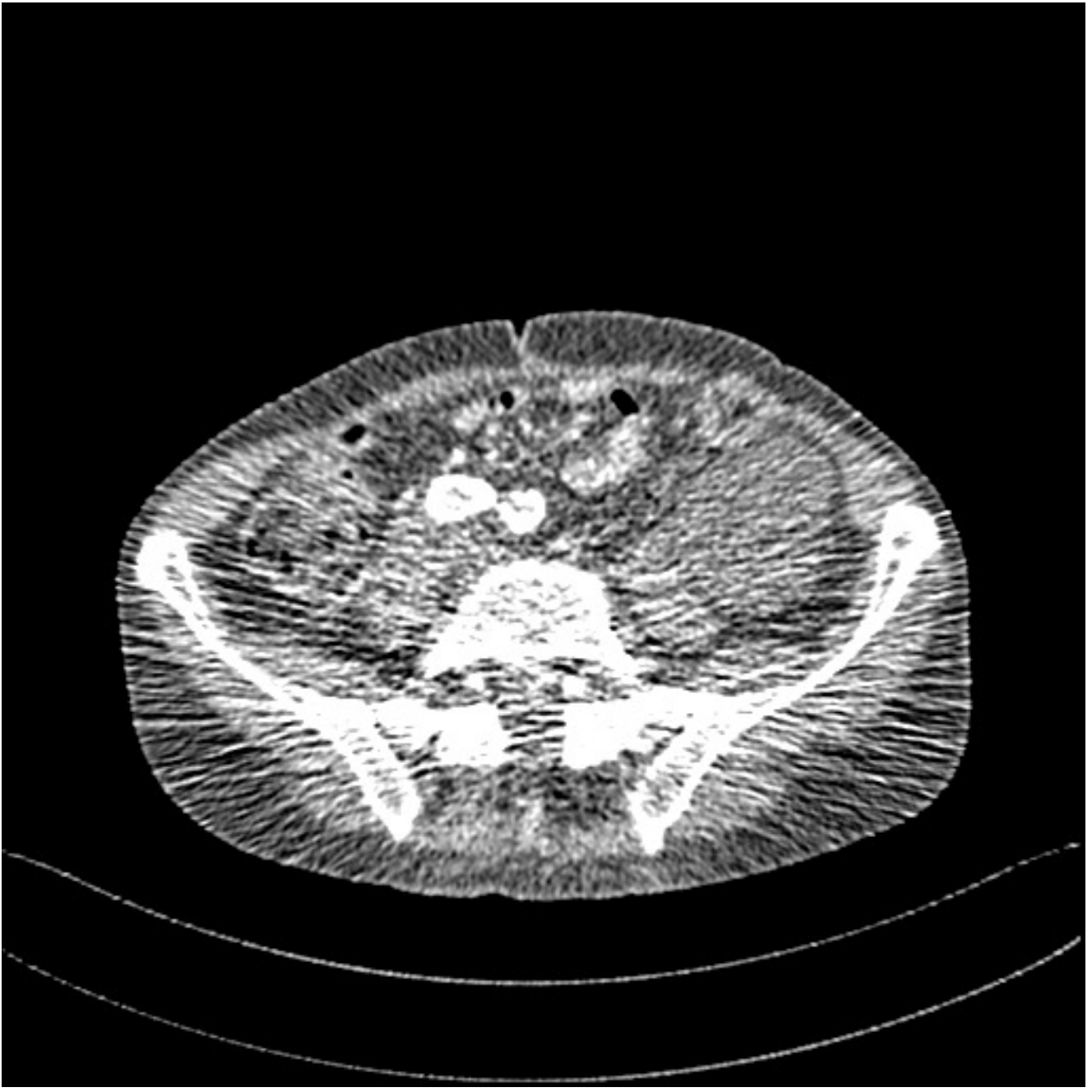}	};
	\spy [green, draw, height = 0.9cm, width = 0.9cm, magnification = 2,
	connect spies] on (1.35,-0.05) in node [left] at (2.00,-1.60);
	\spy [green, draw, height = 0.9cm, width = 0.9cm, magnification = 3,
	connect spies] on (-0.25,-0.05) in node [left] at (-1.12,-1.60);
	\spy [green, draw, height = 1.2cm, width = 1.2cm, magnification = 1.5,
	connect spies] on (-1.40,-0.20) in node [left] at (-0.85,1.45);	
	\draw[line width=1pt, ultra thin, -latex, red] (-0.15,-0.10) -- node[xshift=-0.05cm,yshift=-0.09cm] {\tiny{}} (-0.25,-0.00);	
	\end{tikzpicture}
	\put(-86,110){ \color{white}{\bf \small{RMSE:65.3}}}
	\put(-86,100){ \color{white}{\bf \small{SSIM:0.461}}}
	\put(-75,7){ \color{white}{\bf \small{FBP}}} 
	\\
	\vspace{-0.10in}
	\begin{tikzpicture}
	[spy using outlines={rectangle,green,magnification=2,size=9mm, connect spies}]
	\node {\includegraphics[width=0.23\textwidth]{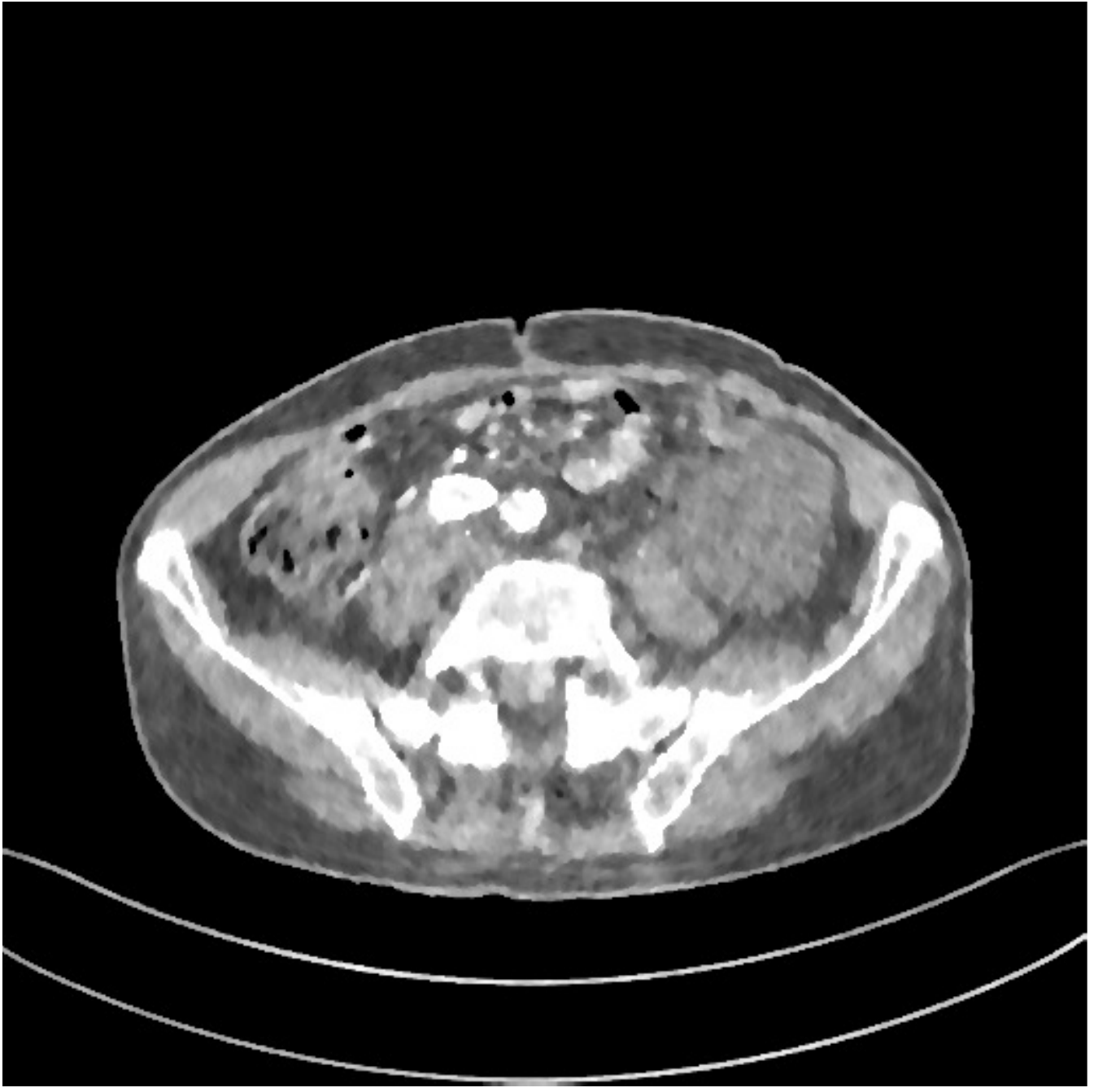}	};
	\spy [green, draw, height = 0.9cm, width = 0.9cm, magnification = 2,
	connect spies] on (1.35,-0.05) in node [left] at (2.00,-1.60);
	\spy [green, draw, height = 0.9cm, width = 0.9cm, magnification = 3,
	connect spies] on (-0.25,-0.05) in node [left] at (-1.12,-1.60);
	\spy [green, draw, height = 1.2cm, width = 1.2cm, magnification = 1.5,
	connect spies] on (-1.40,-0.20) in node [left] at (-0.85,1.45);
	\draw[line width=1pt, ultra thin, -latex, red] (-0.15,-0.10) -- node[xshift=-0.05cm,yshift=-0.09cm] {\tiny{}} (-0.25,-0.00);	
	\end{tikzpicture}
	\put(-86,110){ \color{white}{\bf \small{RMSE:34.3}}}
	\put(-86,100){ \color{white}{\bf \small{SSIM:0.778}}}
	\put(-70,7){ \color{white}{\bf \small{EP}}}
	\hspace{-0.12in}
	\begin{tikzpicture}
	[spy using outlines={rectangle,green,magnification=2,size=9mm, connect spies}]
	\node {\includegraphics[width=0.23\textwidth]{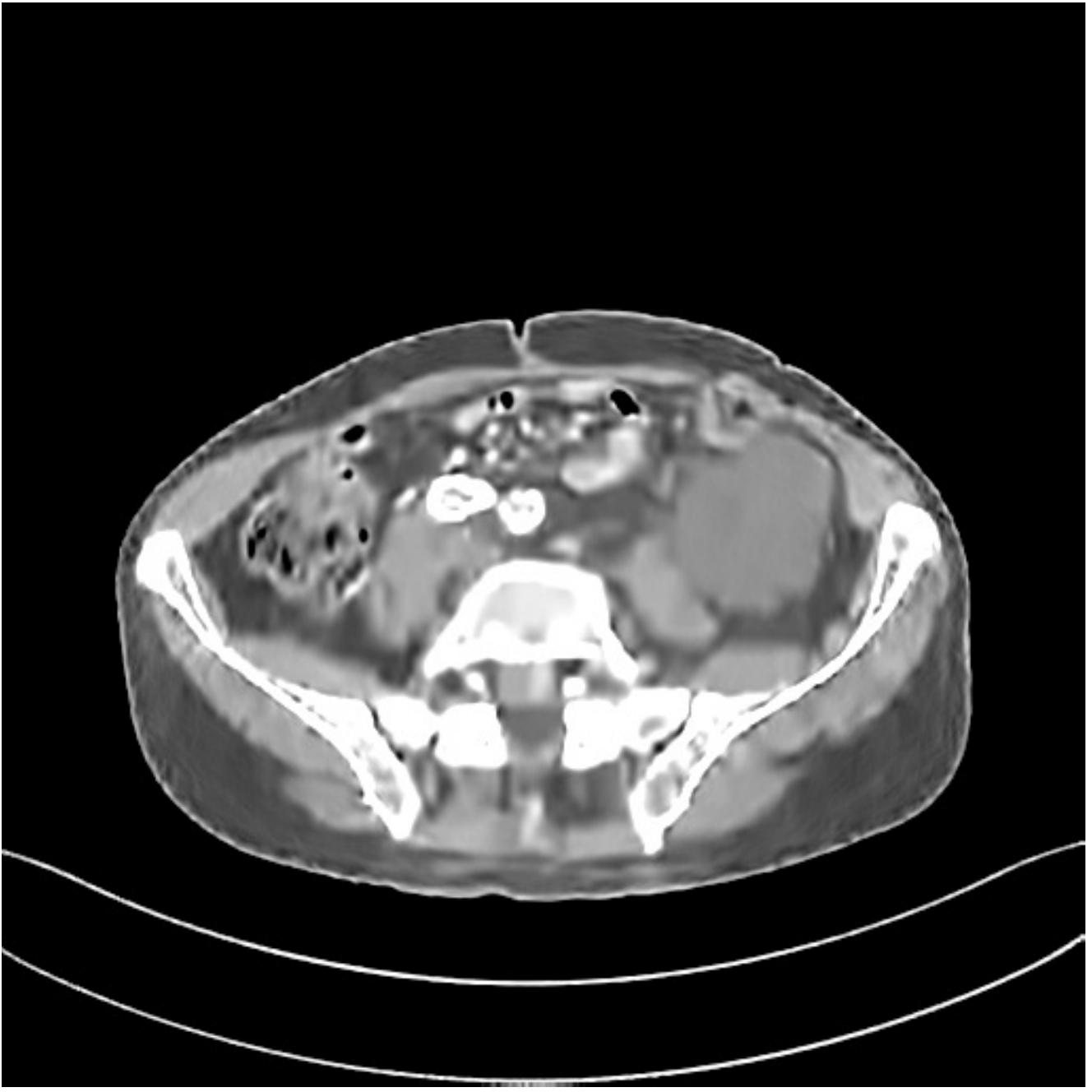}	};
	\spy [green, draw, height = 0.9cm, width = 0.9cm, magnification = 2,
	connect spies] on (1.35,-0.05) in node [left] at (2.00,-1.60);
	\spy [green, draw, height = 0.9cm, width = 0.9cm, magnification = 3,
	connect spies] on (-0.25,-0.05) in node [left] at (-1.12,-1.60);
	\spy [green, draw, height = 1.2cm, width = 1.2cm, magnification = 1.5,
	connect spies] on (-1.40,-0.20) in node [left] at (-0.85,1.45);	
	\draw[line width=1pt, ultra thin, -latex, red] (-0.15,-0.10) -- node[xshift=-0.05cm,yshift=-0.09cm] {\tiny{}} (-0.25,-0.00);
	\end{tikzpicture}
	\put(-86,110){ \color{white}{\bf \small{RMSE:26.0}}}
	\put(-86,100){ \color{white}{\bf \small{SSIM:0.772}}}
	\put(-82,7){ \color{white}{\bf \small{MRST2}}}
	\\
	\vspace{-0.10in}
	\begin{tikzpicture}
	[spy using outlines={rectangle,green,magnification=2,size=9mm, connect spies}]
	\node {\includegraphics[width=0.23\textwidth]{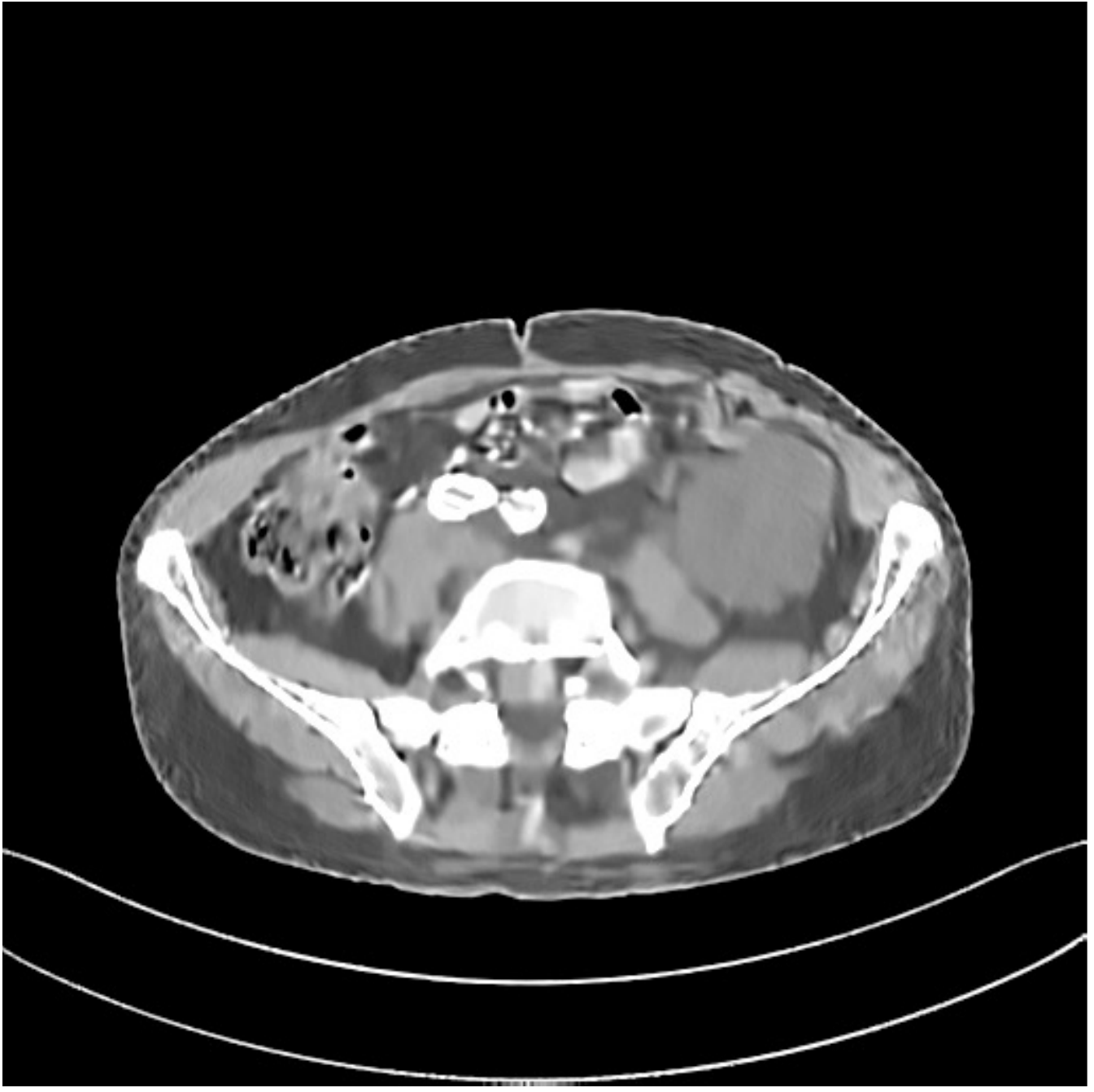}	};
	\spy [green, draw, height = 0.9cm, width = 0.9cm, magnification = 2,
	connect spies] on (1.35,-0.05) in node [left] at (2.00,-1.60);
	\spy [green, draw, height = 0.9cm, width = 0.9cm, magnification = 3,
	connect spies] on (-0.25,-0.05) in node [left] at (-1.12,-1.60);
	\spy [green, draw, height = 1.2cm, width = 1.2cm, magnification = 1.5,
	connect spies] on (-1.40,-0.20) in node [left] at (-0.85,1.45);	
	\draw[line width=1pt, ultra thin, -latex, red] (-0.15,-0.10) -- node[xshift=-0.05cm,yshift=-0.09cm] {\tiny{}} (-0.25,-0.00);		
	\end{tikzpicture}
	\put(-86,110){ \color{white}{\bf \small{RMSE:26.5}}}
	\put(-86,100){ \color{white}{\bf \small{SSIM:0.761}}}
	\put(-82,7){ \color{white}{\bf \small{ULTRA}}}
	\hspace{-0.12in}
	\begin{tikzpicture}
	[spy using outlines={rectangle,green,magnification=2,size=9mm, connect spies}]
	\node {\includegraphics[width=0.23\textwidth]{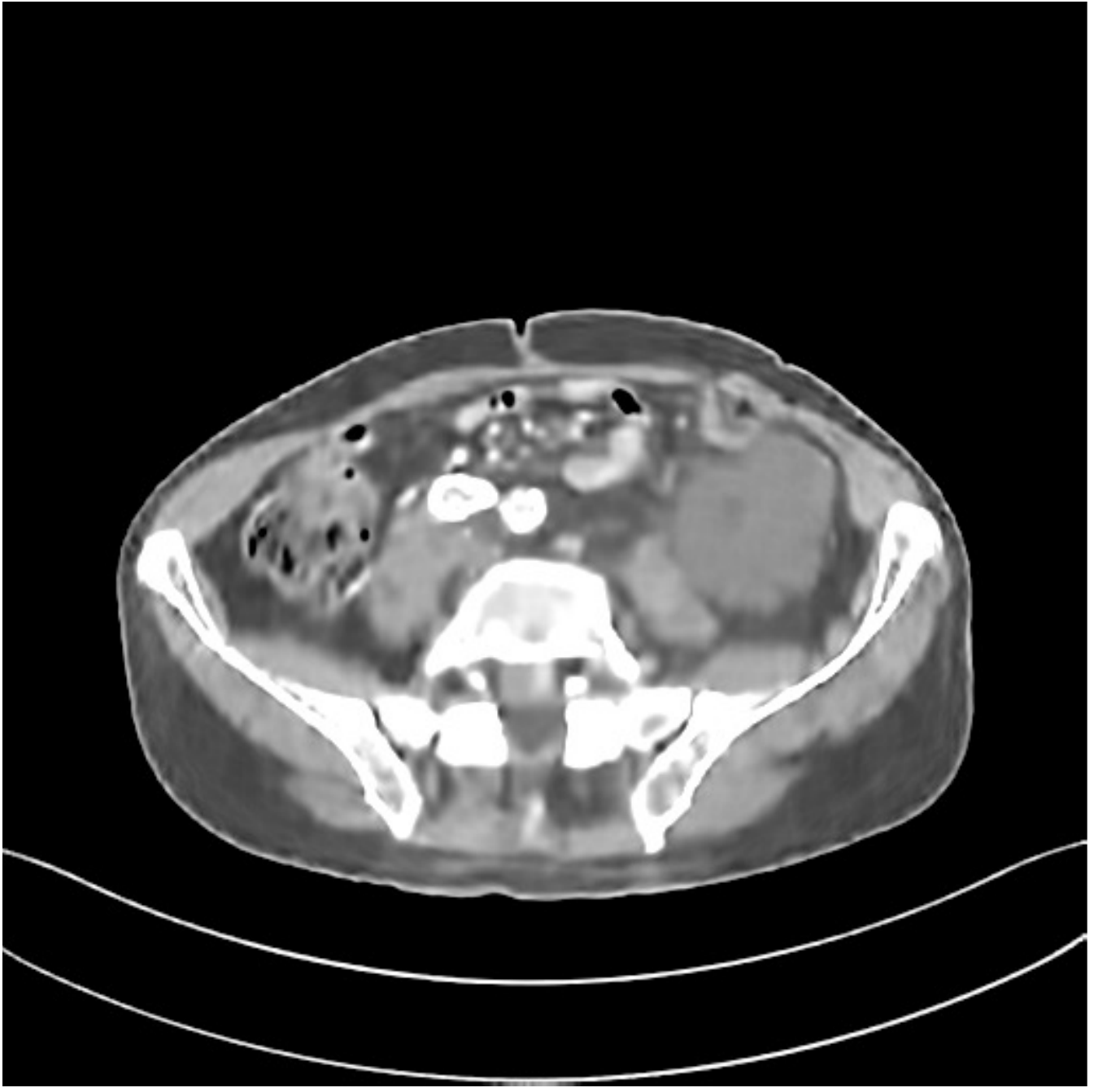}	};
	\spy [green, draw, height = 0.9cm, width = 0.9cm, magnification = 2,
	connect spies] on (1.35,-0.05) in node [left] at (2.00,-1.60);
	\spy [green, draw, height = 0.9cm, width = 0.9cm, magnification = 3,
	connect spies] on (-0.25,-0.05) in node [left] at (-1.12,-1.60);
	\spy [green, draw, height = 1.2cm, width = 1.2cm, magnification = 1.5,
	connect spies] on (-1.40,-0.20) in node [left] at (-0.85,1.45);	
	\draw[line width=1pt, ultra thin, -latex, red] (-0.15,-0.10) -- node[xshift=-0.05cm,yshift=-0.09cm] {\tiny{}} (-0.25,-0.00);		
	\end{tikzpicture}
	\put(-86,110){ \color{green}{\bf \small{RMSE:25.4}}}
	\put(-86,100){ \color{green}{\bf \small{SSIM:0.796}}}
	\put(-82,7){ \color{white}{\bf \small{MCST2}}}
	\vspace{-0.15in}
	\caption{Comparison of reconstructions of one slice from the Mayo Clinic data  with the FBP, PWLS-EP, PWLS-MRST2, PWLS-ULTRA, and PWLS-MCST2 schemes, respectively, at incident photon intensity $I_0=1\times 10^{4}$. The display window is [800, 1200] HU.
	}
	\vspace{-0.20in}
	\label{fig:recon_Mayo_L506}
\end{figure}

\vspace{-0.05in}
We compare our method with the conventional FBP and PWLS-EP algorithms. PWLS-MRST2 \cite{zheng:20:trs} and PWLS-ULTRA \cite{zheng:18:pua} are also included to verify the usefulness of the proposed MCST2 model. We set $\beta = 2^{15.5}$ for PWLS-EP and ran $1000$ iterations of the algorithm to ensure convergence. We ran $1500$ iterations for the other iterative algorithms. The parameters for the XCAT phantom and Mayo Clinic data experiments with the three transform learning-based methods are as follows: $(\beta,\gamma_1,\gamma_2)$ $=$ $(7\times 10^4,30,10)$ and $(2\times 10^4,30,12)$ for PWLS-MRST2; $(\beta,\gamma)$ $=$ $(2\times 10^5,20)$ and $(5\times 10^4,20)$ for PWLS-ULTRA; $(\beta,\gamma_1,\gamma_2)$ $=$ $(1.5\times 10^5,20,5)$ and $(4.5\times 10^4,25,5)$ for PWLS-MCST2. Fig.~\ref{fig:recon_XCAT_slice48} and~\ref{fig:recon_Mayo_L506} show the reconstructions of slices of the XCAT phantom and Mayo Clinic data, respectively. Apart from significantly outperforming the traditional FBP and PWLS-EP methods, the proposed PWLS-MCST2 method performs the best in terms of both RMSE and SSIM compared to the recent MRST2 and ULTRA schemes. Furthermore, PWLS-MCST2 improves the image reconstruction quality by removing more notorious artifacts in the margin regions and preserving critical details in the central region.
\vspace{-0.05in}

\vspace{-0.15in}
\section{Conclusion}
\vspace{-0.10in}
This paper proposes learning a two-layer clustering-based sparsifying transform model (MCST2) for CT images, wherein both the image data and feature (transform residual) maps are divided into multiple classes, with sparsifying filters learned for each class. We present an exact block coordinate descent algorithm to train the MCST2 model from limited unpaired (clean) training data. Our experimental results with simulated XCAT phantom and Mayo Clinic data illustrate that the PWLS approach incorporating the learned MCST2 regularizer outperforms recently proposed MRST2 and ULTRA models. It also provides a significant improvement compared to conventional FBP and PWLS-EP methods. Future work will incorporate and explore deeper MCST models as well as other imaging applications.
\vspace{-0.15in}
%\clearpage
\section{Compliance with Ethical Standards}
\vspace{-0.10in}
This research study was conducted retrospectively using human subject data made available for open access at \url{https://doi.org/10.7937/9npb-2637}. Ethical approval was not required as confirmed by the license attached with the open access data.

\section{Acknowledgments}
\vspace{-0.10in}
The authors thank Dr. Cynthia McCollough, the Mayo Clinic, the American Association of Physicists in Medicine, and the National Institute of Biomedical Imaging and Bioengineering for providing the Mayo Clinic data.

This work was supported in part by the National Natural Science Foundation of China under Grant 61501292. \textsl{ (Corresponding author: Yong Long. Email: yong.long@sjtu.edu.cn)}

%\vfill
%\pagebreak

\bibliographystyle{IEEEbib}
\bibliography{refs}

\begin{thebibliography}{10}

\bibitem{feldkamp:84:pcb}
L.~A. Feldkamp, L.~C. Davis, and J.~W. Kress,
\newblock ``Practical cone beam algorithm,''
\newblock {\em {J. Opt. Soc. Am. A}}, vol. 1, no. 6, pp. {612--619}, June 1984.

\bibitem{fessler:00:sir}
J.~A. Fessler,
\newblock ``Statistical image reconstruction methods for transmission
  tomography,''
\newblock in {\em Handbook of Medical Imaging, Volume 2. Medical Image
  Processing and Analysis}, M.~Sonka and J.~Michael Fitzpatrick, Eds., pp.
  1--70. Proc. SPIE, Bellingham, 2000.

\bibitem{cho:15:rdf}
J.~H. Cho and J.~A. Fessler,
\newblock ``Regularization designs for uniform spatial resolution and noise
  properties in statistical image reconstruction for {3D} {X-ray} {CT},''
\newblock {\em {IEEE Trans. Med. Imag.}}, vol. 34, no. 2, pp. {678--689}, Feb.
  2015.

\bibitem{aharon:06:ksa}
M.~Aharon, M.~Elad, and A.~Bruckstein,
\newblock ``{K-SVD:} an algorithm for designing overcomplete dictionaries for
  sparse representation,''
\newblock {\em IEEE Trans. Sig. Proc.}, vol. 54, no. 11, pp. {4311--4322}, Nov.
  2006.

\bibitem{rubinstein:13:aka}
R.~Rubinstein, T.~Peleg, and M.~Elad,
\newblock ``Analysis {K-SVD}: A dictionary-learning algorithm for the analysis
  sparse model,''
\newblock {\em IEEE Trans. Sig. Proc.}, vol. 61, no. 3, pp. 661--677, Feb.
  2013.

\bibitem{ravishankar:13:lst}
S.~Ravishankar and Y.~Bresler,
\newblock ``Learning sparsifying transforms,''
\newblock {\em IEEE Trans. Sig. Proc.}, vol. 61, no. 5, pp. {1072--1086}, Mar.
  2013.

\bibitem{zheng:18:pua}
X.~Zheng, S.~Ravishankar, Y.~Long, and J.~A. Fessler,
\newblock ``{PWLS-ULTRA}: An efficient clustering and learning-based approach
  for low-dose {3D CT} image reconstruction,''
\newblock {\em IEEE Trans. Med. Imag.}, vol. 37, no. 6, pp. 1498--1510, June
  2018.

\bibitem{ravishankar:18:lml}
S.~Ravishankar and B.~Wohlberg,
\newblock ``Learning multi-layer transform models,''
\newblock in {\em {Allerton Conf. on Comm., Control, and Computing}}, 2018, pp.
  {160--165}.

\bibitem{yang:20:lmr}
X.~{Yang}, X.~{Zheng}, Y.~{Long}, and S.~{Ravishankar},
\newblock ``Learned multi-layer residual sparsifying transform model for
  low-dose {CT} reconstruction,''
\newblock in {\em The 6th International Conference on Image Formation in X-Ray
  Computed Tomography}, 2020, pp. 228--231.

\bibitem{yang:20:mars}
X.~{Yang}, Y.~{Long}, and S.~{Ravishankar},
\newblock ``Multi-layer residual sparsifying transform ({MARS}) model for
  low-dose {CT} image reconstruction,'' 2020,
\newblock Online: \url{https://arxiv.org/abs/2010.06144}.

\bibitem{ravishankar:15:lst}
S.~Ravishankar and Y.~Bresler,
\newblock ``{$\ell_0$} sparsifying transform learning with efficient optimal
  updates and convergence guarantees,''
\newblock {\em IEEE Trans. Sig. Proc.}, vol. 63, no. 9, pp. {2389--2404}, May
  2015.

\bibitem{nien:16:rla}
H.~Nien and J.~A. Fessler,
\newblock ``Relaxed linearized algorithms for faster {X-ray} {CT} image
  reconstruction,''
\newblock {\em IEEE Trans. Med. Imag.}, vol. 35, no. 4, pp. 1090--1098, Apr.
  2016.

\bibitem{ding:16:mmp}
Q.~Ding, Y.~Long, X.~Zhang, and J.~A. Fessler,
\newblock ``Modeling mixed {Poisson-Gaussian} noise in statistical image
  reconstruction for {X-ray} {CT},''
\newblock in {\em Proc. 4th Intl. Mtg. on image formation in X-ray CT}, 2016,
  pp. {399--402}.

\bibitem{zheng:20:trs}
X.~{Zheng}, S.~{Ravishankar}, Y.~{Long}, M.~L. {Klasky}, and B.~{Wohlberg},
\newblock ``Two-layer residual sparsifying transform learning for image
  reconstruction,''
\newblock in {\em 2020 IEEE 17th International Symposium on Biomedical Imaging
  (ISBI)}, 2020, pp. 174--177.

\end{thebibliography}

\end{document}